\documentclass[10pt,a4paper]{article}
\usepackage[a4paper, left=2cm, right=2cm, top=2cm]{geometry}

\usepackage[utf8]{inputenc}
\usepackage{amsmath, amsthm, amssymb, amsfonts}     
\usepackage{mathrsfs}                               
\usepackage{accents}                                
\usepackage{natbib}                                 
\usepackage{hyperref}                               
\usepackage{doi}                                    
\usepackage{authblk}                                
\usepackage{siunitx}                                
\usepackage{booktabs}                               
\usepackage{ulem}                                   
\usepackage{nomencl}                                
\makenomenclature                                   
\usepackage[linesnumbered,ruled,vlined]{algorithm2e}
\SetCommentSty{mycommfont}
\usepackage{subfigure}

\usepackage{bm}                     

\usepackage{graphicx}
\usepackage{tikz}
\usepackage{pgfplots}

\usetikzlibrary{patterns}

\usetikzlibrary{decorations.pathmorphing,arrows,intersections,arrows.meta,angles,quotes}

\pgfplotsset{
	kurze Legende/.style={
		legend image code/.code={
			\draw[##1,mark repeat=2,line width=0.6pt]
			plot coordinates {
				(0cm,0cm)
				(0.3cm,0cm)
			};
		}
	}
}

\pgfplotsset{
	compat = newest,
	scale only axis, 
	max space between ticks = 50pt,
	ticklabel style = {font=\footnotesize},
	legend style =  {font=\footnotesize},
	grid = major,
	grid style = {dotted},
	legend columns=1, 
	xtick pos=left,
	ytick pos=left
}

\pgfplotsset{select coords between index/.style 2 args={
		x filter/.code={
			\ifnum\coordindex<#1\fi
			\ifnum\coordindex>#2\fi
		}
}}

\definecolor{color1}{HTML}{0060AD} 
\definecolor{color2}{HTML}{FF4500} 
\definecolor{color3}{HTML}{FFA500} 
\definecolor{color4}{HTML}{006400} 
\definecolor{color5}{HTML}{9400D3} 
\definecolor{color6}{HTML}{800000} 
\definecolor{color7}{HTML}{000000} 
\definecolor{color8}{HTML}{0000FF} 
\definecolor{color9}{HTML}{FF0000} 
\definecolor{mycolor_blue}{RGB}{66,124,161}
\definecolor{mycolor_grey}{RGB}{198,198,198} 

\tikzstyle{line1} = [color=color7,semithick] 
\tikzstyle{line2} = [color=color2,densely dotted,semithick]
\tikzstyle{line3} = [color=color1,densely dashed,semithick]
\tikzstyle{line4} = [color=color5,dash dot,semithick]
\tikzstyle{line5} = [color=color4,dash dot dot,semithick]
\tikzstyle{line6} = [color=color6,semithick]
\tikzstyle{line7} = [color=color3,densely dotted,semithick]
\tikzstyle{line8} = [color=color8,densely dashed,semithick]

\tikzstyle{mark1} = [color=color7,mark=x,mark size=2pt,mark options=solid,semithick] 
\tikzstyle{mark2} = [color=color2,mark=o,mark size=2pt,mark options=solid,semithick]
\tikzstyle{mark3} = [color=color1,mark=*,mark size=2pt,mark options=solid,semithick]
\tikzstyle{mark4} = [color=color5,mark=triangle,mark size=2pt,mark options=solid,semithick]
\tikzstyle{mark5} = [color=color4,mark=square,mark size=2pt,mark options=solid,semithick]
\tikzstyle{mark6} = [color=color6,mark=o,mark size=2pt,mark options=solid,semithick]
\tikzstyle{mark7} = [color=color8,mark=*,mark size=2pt,mark options=solid,semithick]
\tikzstyle{mark8} = [color=color9,mark=triangle,mark size=2pt,mark options=solid,semithick]

\usetikzlibrary{external}
\title{Rapid Aerodynamic Assessment of Flettner Rotor Installations Using an Inviscid CFD Approach}
\author[]{Niklas K\"uhl\thanks{kuehl@hsva.de}}

\affil[]{Hamburg Ship Model Basin, Bramfelder Strasse 164, D-22305 Hamburg, Germany}

\begin{document}

\providetoggle{tikzExternal}
\settoggle{tikzExternal}{true}
\settoggle{tikzExternal}{false}

\maketitle

\begin{abstract}

This paper introduces an inviscid Computational Fluid Dynamics (CFD) approach for the rapid aerodynamic assessment of Flettner rotor systems on ships. The method relies on the Euler equations combined with a dynamic momentum source term to enforce rotor circulation. By avoiding near-wall refinement and relaxing time-step constraints, the approach significantly reduces computational effort, making it particularly suitable for early-stage design tasks such as parametric studies and design space exploration.

Validation against potential flow theory and viscous reference simulations shows that the method captures lift-induced forces and overall aerodynamic trends reliably. The level of agreement with viscous reference data depends on the operating conditions and numerical setup. While moderate deviations are observed for lower spinning ratios and dissipative convection schemes, larger discrepancies occur at higher spinning ratios and with low-diffusion schemes, particularly in the prediction of drag and peak lift.

Three-dimensional simulations, including idealized wind tunnel setups, rotor–rotor interactions, and full-scale ship applications at Reynolds numbers up to $\mathrm{Re}_\mathrm{L} = 10^8$, demonstrate that the method provides consistent qualitative trends and robust force estimates at a fraction of the computational cost of viscous CFD. This makes the approach well-suited as a fast screening tool in early design phases, where large parameter spaces must be evaluated efficiently.

\end{abstract}

\begin{flushleft}
\end{flushleft}

\section{Introduction}

The need for sustainable shipping has become increasingly important in recent years as the maritime industry strives to reduce greenhouse gas emissions and improve ship efficiency. One promising approach is utilizing Wind Assisted Propulsion (WASP). In this context, Flettner rotors are a prominent and highly relevant example, whose technology dates back roughly one century (\cite{greenspan1969theory, seifert2012review}) but is experiencing a renaissance in the modern shipping industry due to new environmental regulations and technological advances. Flettner rotors offer a promising opportunity to utilize their aerodynamic lift, thereby ultimately supporting a ship's propulsion to reduce its overall fuel consumption. Therefore, studying the aerodynamics and flow conditions around these rotors is crucial to optimizing their performance.

The fluid-dynamic design and optimization of shipbuilding components (e.g., hull, rudder, propulsion unit) involve fluid-physical processes at different spatial and temporal scales. Simultaneous consideration of all systems (e.g., resistance and propulsion or sea-keeping and maneuverability) is challenging, so a sequential approach is usually pursued instead. Especially in early design phases, tools with short response times are utilized, and a reduced prediction quality due to their model simplifications is accepted. In this context, inviscid Computational Fluid Dynamics (CFD) methods are often used, providing a reasonable compromise between solution quality and time to solution, e.g., \cite{john1995computational, ferziger2012computational}. More complex calculations are carried out in a later design phase. A classic example is the hydrodynamic ship-line design, which explores a comparatively large design space in the initial design stage using low-effort flow-potential-based wave-resistance approaches. More sophisticated, viscous prediction methods are only applied to selected, pre-optimized design configurations. The outlined process chain may differ slightly from case to case, but it is a common industrial practice in hydrodynamic design.

Driven by the desire to reduce ship emissions, the demand for WASP is increasing. The increased fuel saving through, e.g., Flettner rotors, poses new challenges for the design process. At the same time, a further component must be considered that interacts with the ship's superstructure and must be suitably optimized. To realistically evaluate the performance of Flettner rotors, precise CFD analyses are required that should follow at least a statistical description of flow turbulence, cf. International Towing Tank \cite{ittc2021aero}, or \cite{menter1994two, samareh1999status}. In practice, mainly non-zonal hybrid filtered/averaged methods for modeling flow turbulence have been established, which, however, are associated with considerable numerical effort (e.g., \cite{menter2009review, spalart1997comments}).

Recent investigations have focused on validating CFD models for Flettner rotors under varying operational parameters, including spinning ratios and turbulence models, demonstrating solid agreement with experimental benchmarks, e.g., \cite{de2014preliminary, de2015marine, bordogna2020aerodynamics, kume2022evaluation}. Further research has examined the influence of rotor geometry and flow conditions on lift generation and drag characteristics using high-fidelity CFD approaches, cf. \cite{arief2018design, copuroglu2018analysis, kwon2022parametric, huang2023study, chen2023experimental}. In addition to purely physical, fluid-mechanical investigations, economic considerations are often at the center of research efforts, cf. \cite{talluri2018techno, arabnejad2024zero, hansen2025wind}, usually accompanied by optimization aspects, e.g., \cite{tillig2020design, reche2025predictive, dhainaut2025large}.

Recent developments in data-driven modeling, such as machine learning-based surrogate models, provide alternative approaches for rapid performance prediction once sufficiently high-fidelity training data is available (e.g., \cite{queipo2005surrogate, koziel2013surrogate}). More recent studies demonstrate the increasing relevance of such approaches in aerodynamic applications, where surrogate models can achieve near-real-time prediction speeds and enable efficient design-space exploration, e.g., \cite{yondo2018review, tao2019application}. However, these methods typically require substantial upfront computational effort to generate high-fidelity training datasets and train models, often relying on large-scale CFD simulations (e.g., \cite{peherstorfer2018survey}). Furthermore, their predictive capability may be limited when extrapolating beyond the training domain or when applied to previously unseen configurations, cf. \cite{brunton2020machine}. In contrast, the present method offers a physics-based, directly applicable alternative that does not rely on prior training data and can be readily deployed for new configurations, making it particularly attractive in early design stages or data-scarce scenarios.

This manuscript aims to evaluate the effectiveness and potential benefits of Flettner rotors on ships in an initial design stage using inviscid Eulerian CFD strategies. The presented method uses a volume force approach to model the fluid circulation generated by the Flettner rotor realistically. The presented method follows a classical incompressible Finite-Volume Method with a co-located variable arrangement. The main difference from viscous approaches is the adaptation of the wall boundary conditions from no-slip to slip. In addition to fewer transport equations to solve, the approach allows significantly coarser temporal and spatial discretization, especially in the near-wall region, resulting in reduced numerical effort. The necessary rotor circulation is adjusted by a suitable volume force model in the rotor-adjacent volumes/cells/elements, with the force magnitude adaptively adjusted during runtime to achieve a prescribed rotor rotation rate. To approximate the flow fields governing the above equations for various rotor configurations, we aim to develop an evaluation procedure that minimizes software interfaces (e.g., different geometry and meshing engines) and reuses as many routines of the actual CFD code as possible. Further features of the approach are (a) the ability to predict lift and its induced resistances, (b) rotor-rotor and rotor-superstructure interactions, and (c) the overall possibility to perform a design space exploration while (d) assembling high-parametric input for, e.g., power prediction tools, cf. \cite{reche2021performance, lee2021prediction}

The proposed strategy requires an interface to the simulation software in order to prescribe momentum sources, e.g., via suitable user coding. As a result, the methodology can be transferred to various CFD solvers. Due to the straightforward body-force-based rotor model control, the method is highly automatable, especially given the simple geometry parameterization of Flettner rotors in combination with external grid generators and the consideration of the entire relative wind spectrum, $0^\circ \leq \alpha < 360^\circ$. Hence, the approach allows the systematic investigation of general Flettner rotor configurations, ranging from different dimensions (aspect ratio, end plate to rotor diameter) over position to interaction aspects.

The structure of the manuscript is as follows. The upcoming Sec. \ref{sec:aerodynamic_beginning} gives an overview of the basic model equations, the discrete approximation, and the volume force model. In Sec. \ref{sec:validation_and_calibration}, the method is verified using academic 2D investigations on a rotating circular cylinder and validated against viscous investigations on three further 3D investigations that consider (a) an idealized rotor, (b) a rotor ship, and (c) a rotor-rotor interaction study. Section \ref{sec:application} applies the presented method to the configuration optimization of Flettner rotors on a bulk carrier before the manuscript ends with a summary and outlook in Sec. \ref{sec:conclusion}. The paper defines vectors and tensors in a spatially fixed orthogonal coordinate system, applying Einstein's summation convention to pairs of Latin indices.

\section{Aerodynamic Modeling \& Simulation Approach}
\label{sec:aerodynamic_beginning}

\subsection{Continuous Modeling}
In a conventional, idealized inviscid flow framework, the following set of equations refer to the local balance of mass as well as linear momentum and governs the temporal ($t$) evolution of an incompressible fluid's ($\rho$) velocity $v_\mathrm{i}$ and pressure $p$, viz.
\begin{alignat}{3}
    \frac{\partial v_\mathrm{k}}{\partial x_\mathrm{k}} &= 0 \qquad &&\text{in} \qquad \Omega \label{equ:mass_balance} \, ,\\
    \frac{\partial \rho v_\mathrm{i}}{\partial t} + \frac{\partial v_\mathrm{k} \rho v_\mathrm{i}}{\partial x_\mathrm{k}} + \frac{\partial p }{\partial x_\mathrm{i}} - b_\mathrm{i} &= 0 \qquad &&\text{in} \qquad \Omega \, , \label{equ:momentum_balance}
\end{alignat}
where $b_\mathrm{i}$ refers to a generic body force. Boundary conditions are based on a numerical wind tunnel in which the ship model to be investigated is placed in the center and separated on the waterline by a symmetry plane $\Gamma^\text{symm}$. In contrast to viscous investigations, a slip wall boundary condition is assumed along the ship's superstructure $\Gamma^\text{wall}$, which suppresses the tangential momentum transport and allows for normal, pressure-induced stresses only. As a consequence, the paper's wall and symmetry boundary condition formally coincide. The apparent wind is prescribed along the outer vertical $\Gamma^\text{horizon}$ boundaries and ambient pressure values are enforced at the domain's upper end $\Gamma^\text{top}$. The boundary conditions are briefly noted below:
\begin{alignat}{4}
    \frac{\partial p}{\partial n} &= 0, &&v_i n_i = 0, \frac{\partial v_i}{\partial n} t_i = 0 \qquad &&\text{on} \qquad \Gamma^\text{wall \& symm} \, , \\
    \frac{\partial p}{\partial n} &= 0, &&v_i = v_i^\mathrm{wind} \qquad &&\text{on} \qquad \Gamma^\text{horizon} \, , \\
    p &= p^\mathrm{ambient}, \quad  &&\frac{\partial v_i}{\partial n} = 0 \qquad &&\text{on} \qquad \Gamma^\text{top} \, .
\end{alignat}
Initial conditions follow a constant apparent-wind and ambient pressure field, i.e.,\begin{align}
    v_i = v_i^\mathrm{wind}, \quad p  = p^\mathrm{ambient}  \qquad \text{at} \qquad t=0 \, .
\end{align}

The fundamental modeling idea assumes that the inviscid method should be able to predict lift effects satisfactorily. The method should provide a solid possibility for reasonably fast predictions and estimations of different Flettner rotor configurations. Due to the neglected viscous effects, drag effects should only be insufficiently represented. However, the paper's investigations are carried out on full-scale ships, so the underlying Reynolds numbers are comparatively large, in the order of $\mathrm{Re}_\mathrm{L} = \mathcal{O}(10^8)$, and the flow is firmly situated in the high-Reynolds-number regime.

The inviscid process only allows a normal, pressure-based momentum transfer between the wall and the fluid, not a tangential, friction-based companion. This poses a challenge to couple the rotor's rotation with the fluid motion, thus ensuring correct fluid circulation in the rotor's vicinity. However, for given rotor speed $N^\mathrm{rotor}$ and geometry (surface coordinates $x_\mathrm{i}^\mathrm{rotor}$ and normal vectors $n_\mathrm{i}^\mathrm{rotor}$ as well as center position $o_\mathrm{i}$ and the rotation vector, e.g., $\omega_i = 2 \pi N^\mathrm{rotor} \delta_\mathrm{i3}$) data, the latter is known a priori, i.e.,
\begin{align}
    \Gamma^\mathrm{rotor} = \int v_\mathrm{i} t_\mathrm{i} \mathrm{d} \Gamma^\mathrm{rotor}
    \qquad \qquad \text{with} \qquad \qquad
    \underline{v} = \underline{\omega}^\mathrm{rotor} \times (\underline{x} - \underline{o})^\mathrm{rotor}, \quad
    \underline{t} = \underline{n}^\mathrm{rotor} \times \underline{e}_3 \label{eqn:rotor_circulation}
\end{align}
where $v_\mathrm{i} t_\mathrm{i}$ refers to the rotor's circumferential velocity component. The example assumes a rotor axis aligned with the third $\underline{e}_\mathrm{3}$ spatial direction. Our approach aims at an equal rotor wall-adjacent fluid velocity, i.e., $v_i^\mathrm{fluid} \to v_i^\mathrm{rotor}$. The fluid motion is triggered by an appropriate body force aligned with the rotor's circumferential motion, i.e., $b_i^\mathrm{fluid} \propto v_i^\mathrm{rotor}$. However, we opt for an integral equivalence based on the fluid's circulation instead of striving for a local equivalence that might require volatile body force with pronounced gradients in both circumferential and radial directions. Hence, we iteratively measure the wall-adjacent fluid circulation and utilize the wall-adjacent body force to drive $\Gamma^\mathrm{fluid} = \Gamma^\mathrm{actual} \to \Gamma^\mathrm{target} = \Gamma^\mathrm{rotor}$. Section \ref{sec:rotor_treatment} provides detailed algorithmic aspects regarding this approach, which is fully externally/isolated controllable as long as the flow solver provides a body-force interface -- which is usually the case.

\subsection{Discrete Treatment}
The balance Eqns. \eqref{equ:mass_balance}-\eqref{equ:momentum_balance} are approximated with the flux-based implicit Finite-Volume-Method FreSCo$^+$ based on a collocated variable arrangement, cf. \cite{rung2009challenges, stuck2012adjoint, kuhl2021phd}. A discrete system of equations of size $NP \times NP$ is formulated to solve for the cell-centered flow field variables. This is achieved through an outer fix-point iteration loop indexed by M at a time step T. Each equation corresponds to the balance of a particular control volume P and is used to compute the cell-centered flow variables, i.e., $[ v_i, p]^\mathrm{P,T, M}$. Local approximations are constructed from values sampled at the cell center $\mathrm{P} \in [1, \mathrm{NP}]$ as well as the face centers F that separate the control volume around P from its neighboring volumes.

The system of equations for the outer iteration M is solved using the Portable, Extensible Toolkit for Scientific Computation (PetSC, cf. \cite{petsc-web-page}), whereby Conjugate Gradient (CG) and BI-Conjugate Gradient (BICG) approaches with Jacobi preconditioning are utilized for the pressure and momentum equations, respectively. Both inner and outer convergence criteria are around $\mathcal{O}(10^{-2})$.

We aim for a numerical process that (a) introduces no more numerical viscosity than necessary while simultaneously (b) being numerically robust, allowing for rapid inviscid flow predictions. As the equations are advanced in a pseudo-time, temporal changes follow an implicit first-order (backward) Euler approximation, and the rotor's circulation-adjusting body force is approximated as a cell-centered momentum source. The latter is incorporated in the pressure-velocity coupling in line with an appropriate momentum-weighted or enhanced Rhie-Chow interpolation scheme, cf. \cite{kuhl2022discrete}. Spatial gradients are consistently approximated via the Gauss divergence theorem. The most delicate part refers to the approximation of convective fluxes, i.e.,
\begin{align}
    \int \frac{\partial}{\partial x_k} \left[ \frac{\partial \rho v_\mathrm{k} v_\mathrm{i} }{\partial x_\mathrm{k}} \right] \mathrm{d} \Omega
    \approx
    \sum_{F(P)} \left[ \rho v_\mathrm{k} v_\mathrm{i} \Delta \Gamma_\mathrm{k} \right]^\mathrm{F, T, M} &= \left[ \dot{m} \, v_i \right]^\mathrm{F, T, M} \, .
\end{align}
Two concepts that interpolate the velocity vector $v_i^\mathrm{F}$ to the face are considered in this paper: A (a) flux-blending and (b) total variation diminishing (TVD) $\kappa$-scheme. The former blends the momentum flux through a cell's face between Upwind Differencing (UD) and Central Differencing (CD), i.e.,
\begin{align}
    v_i^\mathrm{F} = \beta v_i^\mathrm{F, UD} + (1 - \beta) v_i^\mathrm{F, CD} \, , \label{eqn:flux_blending}
\end{align}
thus allowing for a user-defined control of the UD amount that introduces stabilizing numerical diffusion. The $\kappa$-scheme refers to a generalized implementation of van Leer's (\cite{vanLeer1979towards}) Monotonic Upstream Scheme for Conservation Laws (MUSCL) scheme as a combination of first-order upstream and higher-order approximations characterized by the parameter $\kappa$. A damping function $\psi$ dynamically limits local gradients and thus decreasing the approximation order and the implementation reads
\begin{alignat}{2}
    v_i^\mathrm{F} = v_i^\mathrm{U} + \frac{1}{2} \left( v_i^\mathrm{D} - v_i^\mathrm{U} \right)
    \left[ \frac{1 + \kappa}{2} \psi(r) + \frac{1-\kappa}{2} r \psi(\frac{1}{r}) \right]
    \quad \text{with} \quad
    r = \frac{v_i^\mathrm{U} - v_i^\mathrm{UU}}{v_i^\mathrm{D} - v_i^\mathrm{U}} \, , \label{equ:kappa_scheme}
    \quad
    \psi(r) &= \mathrm{minmod}(1, \omega_2 r) \, ,\\
    \qquad \text{and} \qquad 
    r \psi(\frac{1}{r}) &= \mathrm{minmod}(r,\omega_1) \, ,
\end{alignat}
where $v_i^\mathrm{D}$, $v_i^\mathrm{U}$, and $v_i^\mathrm{UU}$ denote the face's downwind and two upwind values, respectively. The minmod operator is defined as $\mathrm{minmod}(x, \omega y) = \mathrm{sgn}(x) \, \mathrm{max}(0, \mathrm{min}(|x|, \omega \, y \, \mathrm{sgn}(a)))$ and the two additional argument values read $\omega_1 = \mathrm{max}(1, (9(3-\kappa))/(10(1-\kappa)))$ and $\omega_2 = \mathrm{max}(1, (9(3+\kappa))/(10(1+\kappa)))$. The formulation allows for mixing between centralized and upwind biased strategies, e.g., CD and the Quadratic Upstream Interpolation for Convective Kinematics (QUICK) via $0.5 \leq \kappa \leq 1.0$, e.g., $v_i^\mathrm{F} = v_i^\mathrm{F,CD}$ for $\kappa = 1$ or $v_i^\mathrm{F} = v_i^\mathrm{F, QUICK}$ for $\kappa = 0.5$. Further details on the implementation can be found in \cite{stuck2012adjoint, manzke2018development}. 

The considered numerical approaches refer to standard approximation strategies in prominent CFD packages, allowing the transfer of this paper's approaches. Further details on the numerical infrastructure including solver validation or possibilities for design optimization can be found, e.g., \cite{kuhl2021continuous, kuhl2021adjoint_2, loft2023two} or \cite{mueller2021novel, kuhl2022adjoint, radtke2023parameter}.

\subsection{Rotor Circulation Approximation via Wall-Adjacent Momentum Sources}
\label{sec:rotor_treatment}
The relevant algorithmic relationships are outlined below. Algorithm \ref{alg:flow_solver} essentially controls the time integration and declares the globally required variables needed in the later modules. Relevant integral parameters are also determined. Before the time integration starts, the initial and boundary conditions of the aerodynamic scenario are defined. 
\begin{algorithm}[!h]

\SetKwInput{KwInput}{Input}
\SetKwInput{KwOutput}{Output}
\SetKwInput{KwOutputOpt}{Optional Output}
\DontPrintSemicolon

\SetKwFunction{FMain}{flow\_solver}
 
\SetKwProg{Fn}{Module}{:}{\KwRet}
\Fn{\FMain{}}{
    \KwInput{nR \tcp*{Number of rotors}}
    \KwInput{nUpdate \tcp*{Time step count  for body force update}}
    \KwInput{nSwapDir \tcp*{Time step count for rotation direction check}}
    \KwInput{nTurnOff \tcp*{Time step count for turn-off check}}
    \textbf{PUBLIC} iR \tcp*{Rotor counter}
    \textbf{PUBLIC} nB, iB \tcp*{Number of boundary faces and face counter}
    \textbf{PUBLIC} c$_\mathrm{i}$(nR,3), n(nR) \tcp*{Rotational axis and speed per rotor}
    \textbf{PUBLIC} $f_\mathrm{i}$(nR,3) \tcp*{Forces per rotor}
    \textbf{PUBLIC} $\Gamma^\mathrm{tar}$(nR), $\Gamma^\mathrm{is}$(nR), ratio(nR) \tcp*{Target, actual and ratio of flow circulations}

    \vphantom{da} \\
    call init\_flow\_field() \tcp*{Initialize flow field, cf. Alg. \ref{alg:initial_flow_field}}
    call init\_body\_force\_field() \tcp*{Initialize body force field, cf. Alg. \ref{alg:initial_body_force_field}}
    \For{timeStep=1,nTimeSteps} 
    {
        acquire new velocity $v_\mathrm{i}$ and pressure $p$ values\tcp*{Compute new flow field}
        call body\_force\_update() \tcp*{Measure rotor forces and update the body force field, cf. Alg. \ref{alg:body_force_update}}
    }
}

    \caption{User-coding that initializes the rotor wall-adjacent body-force.}
    \label{alg:flow_solver}
\end{algorithm}
Algorithm \ref{alg:initial_flow_field} initializes the actual wind scenario on the basis of a read-in file (windDataIn.dat) which, in addition to the ship's speed, primarily provides the wind speed and the angle of attack as well as data on the atmospheric profile. This process allows a high degree of automation of the investigations to be carried out. The apparent wind vector is defined and initialized with the help of a rotation matrix around the $x_3$-axis. The initial pressure field corresponds to the constant ambient pressure. 
\begin{algorithm}[!h]

\SetKwInput{KwInput}{Input}
\SetKwInput{KwOutput}{Output}
\SetKwInput{KwOutputOpt}{Optional Output}
\DontPrintSemicolon

\SetKwFunction{FMain}{init\_flow\_field}
 
\SetKwProg{Fn}{Module}{:}{\KwRet}
\Fn{\FMain{}}{
    \KwInput{windDataIn.dat \tcp*{True wind data file}}
    \textbf{PRIVATE} $h^\mathrm{wind}$ \tcp*{Reference wind profile height [m]}
    \textbf{PRIVATE} $v^\mathrm{wind}(3)$ \tcp*{Reference wind magnitude [m/s] at $h^\mathrm{wind}$}
    \textbf{PRIVATE} $v^\mathrm{ship}(3)$ \tcp*{Ship speed [m/s]}
    \textbf{PRIVATE} $\alpha^\mathrm{wind}$ \tcp*{True wind angle of attack [rad] measured from ship aft}
    \textbf{PRIVATE} mExp \tcp*{Power-law exponent}
    \textbf{PRIVATE} $D_\mathrm{ik}$(3,3) \tcp*{Transformation matrix}
    \textbf{PRIVATE} $p^\mathrm{ambient}$ \tcp*{Ambient pressure [Pa]}
    
    \vphantom{da} \\
    $v_1^\mathrm{ship} \gets \mathrm{read(windDataIn.dat)}$ \tcp*{Read ship speed}
    $(v_1^\mathrm{wind}, \alpha^\mathrm{wind}, h^\mathrm{wind}) \gets \mathrm{read(windDataIn.dat)}$ \tcp*{Read wind speed data}
    $p^\mathrm{ambient} \gets \mathrm{read(windDataIn.dat)}$ \tcp*{Read ambient pressure}

    \vphantom{da} \\
    $D_\mathrm{ik} = 0$; $D_\mathrm{11} = \mathrm{cos}(\alpha^\mathrm{wind})$; $D_\mathrm{21} = \mathrm{sin}(\alpha^\mathrm{wind})$ \tcp*{Fill the rotation matrix}
    $D_\mathrm{12} = -D_\mathrm{21}$; $D_\mathrm{22} = D_\mathrm{11}$; $D_\mathrm{33} = 1$ \tcp*{Fill the rotation matrix}

    $v_1^\mathrm{wind} = v_1^\mathrm{wind} (x_3/h^\mathrm{wind})^\mathrm{mExp}$ \tcp*{Atmospheric wind profile in $x_1$ as function of $x_3$ direction}
    $v_\mathrm{i} = D_\mathrm{ik} v_\mathrm{k}^\mathrm{wind} + v_\mathrm{i}^\mathrm{ship}$ \tcp*{Define initial true wind field and boundary conditions}

    \vphantom{da} \\
    $p = p^\mathrm{ambient}$ \tcp*{Define initial pressure field and boundary conditions}
}

    \caption{User-coding that defines the flow field's initial and boundary conditions.}
    \label{alg:initial_flow_field}
\end{algorithm}
Algorithm \ref{alg:initial_body_force_field} first reads all relevant rotor and speed data from a second input file (rotorDataIn$\_$rotorNumber.dat), whereby any number of rotors with different positions and speeds can be examined. The automatic determination of the horizontal rotor position using the zero and first geometric moment is worth mentioning. The same applies to the determination of the rotor axis. This simplifies the use of the overall process, as now only the rotor speed and a surface indicator are required to determine the surfaces associated with the rotor. Based on the data, the target circulation $\Gamma^\mathrm{target}$ of each rotor is calculated. Also, the momentum source in the first computational cells above the rotor wall is initialized appropriately in circumferential direction.
\begin{algorithm}[!h]

\SetKwInput{KwInput}{Input}
\SetKwInput{KwOutput}{Output}
\SetKwInput{KwOutputOpt}{Optional Output}
\DontPrintSemicolon

\SetKwFunction{FMain}{init\_body\_force\_field}
 
\SetKwProg{Fn}{Module}{:}{\KwRet}
\Fn{\FMain{}}{
    \KwInput{rotorDataIn\_*nR*.dat \tcp*{Number of Rotors and input files}}
    \textbf{PRIVATE} $l$, $m_\mathrm{i}$(3) \tcp*{Geometrical moments}
    \vphantom{da} \\
    $f_\mathrm{i}(:) = 0$, $\Gamma^\mathrm{tar}$$(:) = 0$, ratio$(:) = 0$ \tcp*{Initialize rotor forces, target circulations, and ratios}
    \For{iR=1,nR} 
    {
        n(iR)$ \gets \mathrm{read(rotorDataIn\_*iR*.dat)}$\tcp*{Read rotor rotation rate}
        $l = 0, m_\mathrm{i} = 0$\tcp*{Initialize  geometrical moments}
        \For{iB=1,nB}
        {
            \If{iB belongs to iR}
                {
                   $l = l + \sqrt{\Delta \Gamma_i^\mathrm{iB} \Delta \Gamma_i^\mathrm{iB} }$ \tcp*{Update zeroth geometrical moment}
                   $m_\mathrm{i} = m_\mathrm{i} + \sqrt{\Delta \Gamma_i^\mathrm{iB} \Delta \Gamma_i^\mathrm{iB} } x_\mathrm{i}^\mathrm{iB}$ \tcp*{Update first geometrical moment}
                }
        }
        $o_\mathrm{i} = m_\mathrm{i} / l$ \tcp*{Compute geometric center}
        $\omega_\mathrm{i} = 0; \omega_3 = 2 \pi n(iR)$ \tcp*{Compute rotation vector}
        $c_\mathrm{i}(iR) = \omega_\mathrm{i}/|\omega_\mathrm{i}|$ \tcp*{Compute rotational axis}
        \For{iB=1,nB}
        {
            \If{iB belongs to iR}
                {
                   $v_\mathrm{i}^\mathrm{iB} = \omega_\mathrm{i} \times (x_\mathrm{i}^\mathrm{iB} -o_\mathrm{i})$ \tcp*{Circumferential rotor velocity}
                   $\Gamma^\mathrm{tar}\mathrm{(iR)} = \Gamma^\mathrm{tar}\mathrm{(iR)} + v_\mathrm{i}^\mathrm{iB} (\Delta \Gamma_\mathrm{k}^\mathrm{iB} \times c_\mathrm{k}\mathrm{(iR)})_\mathrm{i}$ \tcp*{Update target circulation}
                   $b_\mathrm{i}^\mathrm{iB}  = \rho v_\mathrm{i}^\mathrm{iB} n(iR) $ \tcp*{Initial guess for momentum source }
                }
        }
    }
}

    \caption{Pseudo-code to initialize the rotor-wall adjacent body-force based on the rotor's rotational rate.}
    \label{alg:initial_body_force_field}
\end{algorithm}
In the last algorithm, the actual control of the Flettner rotors is carried out. For this purpose, the current circulations $\Gamma^\mathrm{actual}$ are calculated every time step. The deviation from the target circulations defined in Alg. \ref{alg:initial_flow_field} is considered every  $\#\mathrm{nUpdate}$ time step, and the ratio scales the volume force appropriately. In the ideal, converged state, the ratio reads one. However, actual investigations, especially at high spinning ratios, show a slight fluctuation around the unit value. Since the averaging of rotor forces is also carried out in Alg. \ref{alg:initial_body_force_field}, two further features have been implemented. If the propulsive force is negative up to a specific number of time steps ($\#\mathrm{nSwapDir}$), the current rotational direction seems to not generate thrust but resistance, so that a single change of the ratio variable reverses the direction of rotation. If the rotor still generates resistance ($\#\mathrm{nTurnOff}$), the rotor is deactivated by setting the ratio variable to zero. Both features can be deactivated by selecting the appropriate variables. Typical values of $\#\mathrm{nSwapDir}$ and $\#\mathrm{nTurnOff}$ are chosen, such that 5-10 equivalent rotor flows are considered.
\begin{algorithm}[!h]

\SetKwInput{KwInput}{Input}
\SetKwInput{KwOutput}{Output}
\SetKwInput{KwOutputOpt}{Optional Output}
\DontPrintSemicolon

\SetKwFunction{FMain}{body\_force\_update}
 
\SetKwProg{Fn}{Module}{:}{\KwRet}
\Fn{\FMain{}}{
    $\Gamma^\mathrm{is}= 0$ \tcp*{Initialize actual circulations}
    \For{iR=1,nR} 
    {
    
    \For{iB=1,nB}
    {
        \If{iB belongs to iR}
            {
                $\Gamma^\mathrm{is}\mathrm{(iR)} = \Gamma^\mathrm{is}\mathrm{(iR)} + v_\mathrm{i}^\mathrm{iB} (\Delta \Gamma_\mathrm{k}^\mathrm{iB} \times c_\mathrm{k}\mathrm{(iR)})_\mathrm{i}$ \tcp*{Update target circulation}
                $f_\mathrm{i}$(iR)$ = f_\mathrm{i}$(iR)$ + p$(iB) $\Delta \Gamma_\mathrm{i}$(iB) \tcp*{Update rotor forces}
            }
    }

    ratio(iR)$ = |\Gamma^\mathrm{tar}\mathrm{(iR)}/\Gamma^\mathrm{ist}\mathrm{(iR)}|$

    \If{mod(timestep,nUpdate) = 0}
        {
            $f_\mathrm{i}$(iR) = $f_\mathrm{i}$(iR)/ nUpdate \tcp*{Compute mean rotor forces}
            IF ( (timestep = nSwapDir) \& ($f_1$(iR) $< -10^{-3}$ ) ) ratio(iR) = -ratio(iR) \tcp*{Swap rotation}
            IF ( (timestep = nTurnOff) \& ($f_1$(iR) $< -10^{-3}$ ) ) ratio(iR) = 0 \tcp*{Turn rotor off}
            \For{iB=1,nB}
            {
            \If{iB belongs to iR}
                {
                    $b_\mathrm{i}$(iB)$ = b_\mathrm{i}$(iB)\,ratio(iR) \tcp*{Update body force magnitude}
                }
            }
        $f_\mathrm{i}$(iR) = 0 \tcp*{Re-initialize rotor forces}
        }
    }
}

    \caption{Pseudo-code to update the rotor-wall adjacent body-force.}
    \label{alg:body_force_update}
\end{algorithm}

\section{Concept Validation and Calibration}
\label{sec:validation_and_calibration}
The proposed concept for predicting the performance of Flettner rotors is carefully validated in the following. It starts with academic flow-potential-based investigations on (a) a two-dimensional circular cylinder and increases in complexity towards (b) one three-dimensional isolated rotor, (c) one rotor operating on a real ship up to (d) two interacting rotors. The inviscid approach is compared against theoretical and high-fidelity numerical reference data.

\subsection{2D Flow Around Rotating Circular Cylinder}
\label{subsec:cirular_cylinder}
\begin{figure}[!ht]
    \centering
    \subfigure[]{
    \iftoggle{tikzExternal}{
	\input{./tikz/01__cylinder_2D/rotating_cylinder.tikz}}{
	\includegraphics{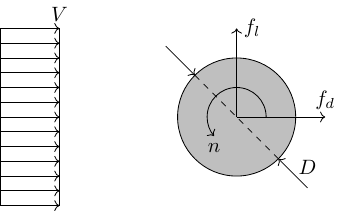}}
    }
    \hspace{1cm}
    \subfigure[]{
    \includegraphics[width=0.45\textwidth]{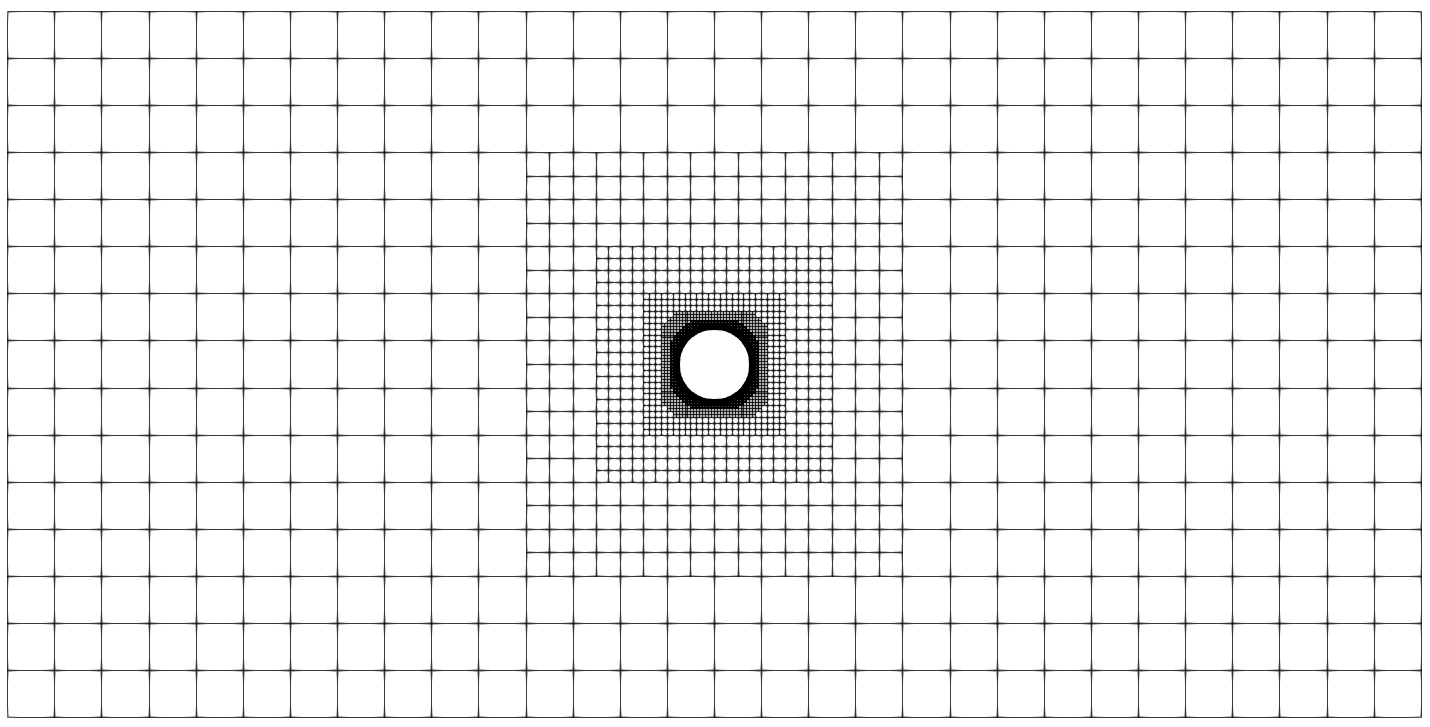}
    }
    \caption{Two dimensional circular cylinder case: Schematic representation of (a) the overall setup and (b) of the utilized numerical grid.}
	\label{fig:rotating_cylinder_2D}
\end{figure}

First, we start with an academic case considering the flow-potential-based investigation of a rotating (rotational speed $n$) circular cylinder of diameter $D$ in a free stream with homogeneous velocity $V$. For the case sketched on the left in Fig. \ref{fig:rotating_cylinder_2D}, analytical solutions for the velocity and pressure field can be derived, providing information about the resulting drag ($f_\mathrm{d}$) and lift ($f_\mathrm{l}$) forces, especially as a function of the rotor's rotational speed. An exemplary, unstructured numerical grid is shown on the right. The grid consists of approximately 9500 finite volumes and is refined towards the cylinder, whereby the cylinder circumference is discretized with roughly 750 cells.

The section's main focus is on the behavior of the rotating cylinder's lift and drag coefficients as a function of the spinning ratio, which are defined for the paper's remainder as follows:
\begin{align}
    c_\mathrm{d} = \frac{2 f_\mathrm{d}}{\rho V^2 D H} \, ,
    \qquad \qquad
    c_\mathrm{l} = \frac{2 f_\mathrm{l}}{\rho V^2 D H}
    \qquad \qquad \text{and} \qquad \qquad
    k = \frac{\pi n D}{V} \, .
\end{align}
This section's benchmark values refer to the ideal, potential-flow-based relation, i.e., $c_\mathrm{l}(k) = 2 \pi k$ and $c_\mathrm{d}(k) = 0$.

\begin{figure}[!htb]
\centering
\iftoggle{tikzExternal}{
\input{./tikz/01__cylinder_2D/exemplary_results.tikz}}{
\includegraphics{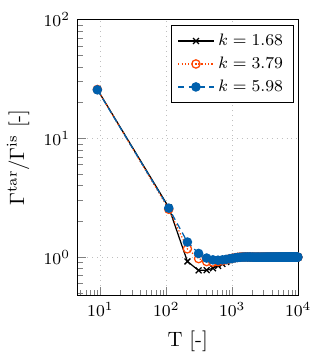}
\includegraphics{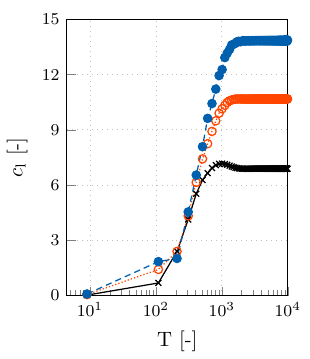}
\includegraphics{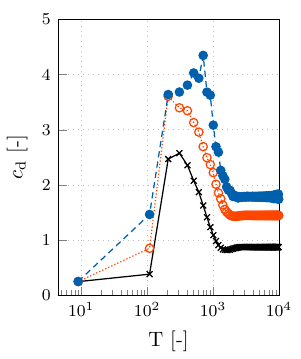}
}
\caption{Two dimensional circular cylinder case: Exemplary development of (left) the ratio of the target/actual flow circulation, (center) the lift coefficient, and (right) the drag coefficient over the simulated time steps.}
\label{fig:rotating_cylinder_exemplary_results}
\end{figure}

Figure \ref{fig:rotating_cylinder_exemplary_results} shows an exemplary response of the proposed dynamic rotor rotation adjustment procedure outlined in Sec. \ref{sec:rotor_treatment} (cf. Algs. \ref{alg:flow_solver} - \ref{alg:body_force_update}) with UD approximation, i.e., $\beta = 1$ in Eqn. \eqref{eqn:flux_blending}. For three spinning ratios 
, the development of the ratio of the target/actual flow circulation (cf. Eqn. \eqref{eqn:rotor_circulation})), the lift coefficient, and the drag coefficient are depicted over the simulated time steps from left to right, respectively. The discrete markings indicate the adjustment of the volume force every 100-time steps, cf. Alg. \ref{alg:body_force_update}. All three circulation ratios start at circulations that are more than one magnitude too small and converge to the targeted unit ratio within the first 1000 time steps. This convergence can also be observed in the lift and drag coefficients, which reach a plateau with a slight delay. As expected, the lift coefficient increases for higher spinning ratios, whereby the lift coefficients are more than an order of magnitude higher than their drag companions. The latter are different from zero, indicating the violation of the inviscid flow assumption. This is due, in particular, to the upwind-biased approximation of the convective term used and the associated introduction of a solid numerical diffusion.

The exemplary investigation is consistently continued for different spinning ratios in the interval $0 \leq k \leq 8$, whereby different aspects of the numerical process, like the resolution of the outer iterative method or different convection schemes, are considered. Temporal and spatial discretization influences have been eliminated in a preceding grid and time step study. The time step is chosen to discretize one equivalent flow based on the free stream velocity with approximately 100-time steps. The numerical responses tend to fluctuate, especially for higher spinning ratios, so results are averaged over the last 5 equivalent flows, i.e., 500-time steps.

\begin{figure}[!htb]
\centering
\iftoggle{tikzExternal}{
\input{./tikz/01__cylinder_2D/uds_cds_blending_ratio.tikz}}{
\includegraphics{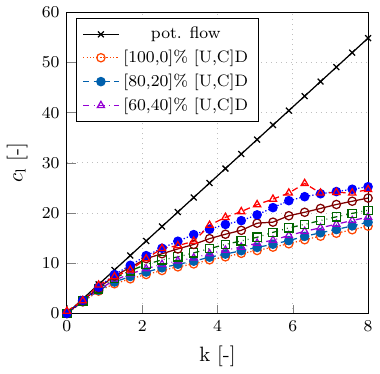}
\includegraphics{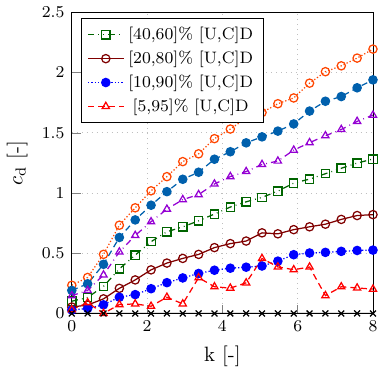}
}
\caption{Two dimensional circular cylinder case: Predicted lift (left) and drag (right) coefficient over the spinning ratio for different Upwind Differencing (UD) and Central Differencing (CD) ratios to approximate the convective momentum transport.}
\label{fig:rotating_cylinder_uds_cds_blendings}
\end{figure}

First, the flux blending approach from Eqn. \eqref{eqn:flux_blending} is examined for seven UD/CD ratios based on the blending parameter $\beta$ such that 100 (0) up to 5 (95) percent upwind (central) differencing are examined. The limit case of 0 (100) percent UD (CD) tends to produce numerical instabilities and is thus not considered. Resulting lift and drag coefficients are shown in Fig. \ref{fig:rotating_cylinder_uds_cds_blendings} on the left and right over the spinning ratio, respectively. Additionally, the reference potential flow results are included.
An increase in the CD/UD ratio leads to a moderate increase in the lift and a reduction in the drag coefficient. While the differences in the lift coefficient become visible from a spinning ratio of $k \geq 1$, the differences in the drag companion are visible across the entire spinning ratio spectrum and are much more pronounced. The results align with theoretical considerations, which dictate a decreased numerical viscosity for reduced upwind proportions.

\begin{figure}[!htb]
\centering
\iftoggle{tikzExternal}{
\input{./tikz/01__cylinder_2D/vary_nIter_lift.tikz}}{
\includegraphics{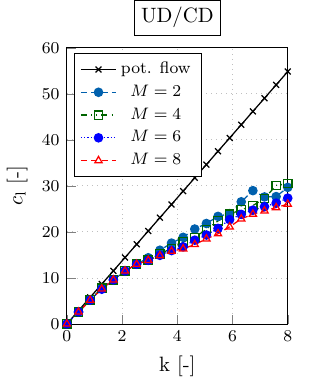}
\includegraphics{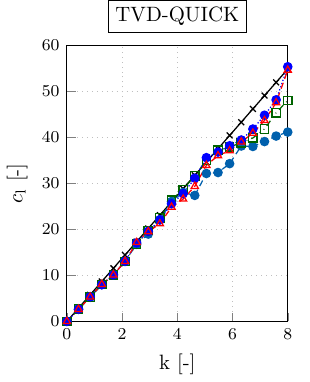}
\includegraphics{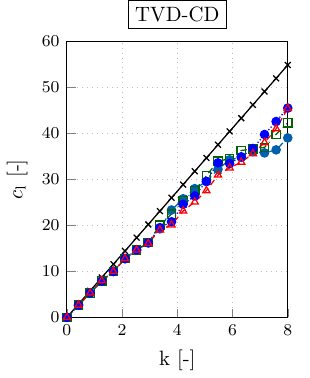}
}
\caption{Two dimensional circular cylinder case: Lift coefficient for different fixed outer iterations over the spinning ratio for (left) a flux blending case with 10/90 percent UD/CD ($\beta=0.1$), (center) a TVD-QUICK approach with $\kappa = 0.5$, and (right) a TVD-CD case with $\kappa = 1.0$.}
\label{fig:rotating_cylinder_vary_nIter_lift}
\end{figure}

\begin{figure}[!htb]
\centering
\iftoggle{tikzExternal}{
\input{./tikz/01__cylinder_2D/vary_nIter_drag.tikz}}{
\includegraphics{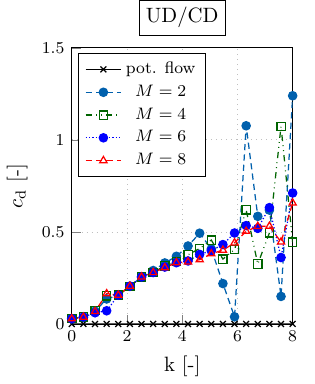}
\includegraphics{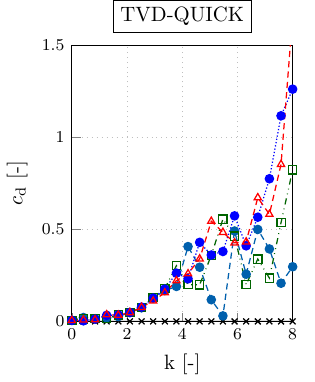}
\includegraphics{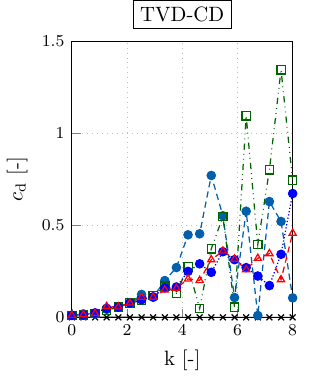}
}
\caption{Two dimensional circular cylinder case: Drag coefficient for different fixed outer iterations over the spinning ratio for (left) a flux blending case with 10/90 percent UD/CD ($\beta=0.1$), (center) a TVD-QUICK approach with $\kappa = 0.5$, and (right) a TVD-CD case with $\kappa = 1.0$.}
\label{fig:rotating_cylinder_vary_nIter_drag}
\end{figure}

In the following, in addition to the 10/90 \% UD/CD ($\beta=0.1$) flux blending case, TVD scenarios with $\kappa=0.5$ and $\kappa=1.0$ are examined, whereby the number of fixed, enforced outer iterations is varied, i.e., $M = [2,4,6,8]$. The investigations are intended to check whether the numerical effort can be reduced by decreasing the convergence per time step while maintaining the quality of the results. The results are shown in Fig. \ref{fig:rotating_cylinder_vary_nIter_lift} for the lift coefficient and in Fig. \ref{fig:rotating_cylinder_vary_nIter_drag} for the drag coefficient, with the flux-blending results on the left, the TVD-QUICK results in the middle, and the TVD-CD results on the right. Minor differences can be identified in the lift coefficient between the investigations at different numbers of outer iterations, which are particularly pronounced at higher spinning ratios. Up to $k=4$, hardly any differences are recognizable in both coefficients. Instead, predictions from the employed convection schemes differ: While the lift coefficients of the flux-blending investigations correspond to the results from Fig. \ref{fig:rotating_cylinder_uds_cds_blendings} and visibly deviate from the flow-potential-based reference results for $k \geq 2$, the TVD investigations align with the reference results for higher spinning ratios. Predictions of the TVD-CD method also deviate from the benchmark results for $k \geq 2$ but still follow the reference trend with a less reduced slope. The TVD-QUICK method's results align mostly perfectly with the reference lift results up to $k \leq 4$ and are still close to $k \leq 6$. The differences between both TVD approaches indicate that the TVD-CD method increasingly switches dynamically to low-order, UD-biased approximations.
Results of the predicted resistance coefficients in Fig. \ref{fig:rotating_cylinder_vary_nIter_drag} allow similar conclusions. Note: Due to considerably less numerical diffusion, the y-axis represents a smaller interval than in Fig. \ref{fig:rotating_cylinder_uds_cds_blendings} (right). The TVD methods generally outperform the flux-blending approach up to $k \leq 4$ with hardly any difference between the results for different numbers of outer iterations. For increased spinning ratios $k \geq 4$, the different studies provide stronger deviating predictions, which are additionally subject to solid fluctuations, that are highest for several (a few) outer iterations in the case of the TVD-based (flux-blending) investigations. Additionally, the increase in the drag coefficient for TVD-CD starts somewhat earlier at $k=1$ than at $k=2$ for TVD-QUICK. However, TVD-CD results feature a smaller slope, resulting in a slightly lower drag coefficients for $k=4$.

\paragraph{A summary in between:} About 10-20 adjustments of the near-cylinder volume force are necessary to achieve the target circulation. Furthermore, the numerical effort can be significantly reduced by not converging reasonably every time step. However, fixed 2-4 outer iterations are sufficient for solid predictions, especially in the technically relevant spinning ratio range up to $k \leq 4$. The closest solutions to the ideal flow-potential-based reference data are predicted by the TVD-QUICK method, where the lift results are remarkably well reproduced up to $k=5$ for the lift coefficient and up to $k=2$ for the drag companion. In contrast, all investigated flux-blending procedures introduce a solid amount of numerical viscosity into the numerical system, resulting in significantly increased drag coefficients and, at the same time, reduced lift coefficients, already at moderate ($k=2$) and especially at increased ($k=4$) spinning ratios. However, compared to the idealized, inviscid flow assumption, this deficiency can be exploited by suitable model calibration for real-world, viscous applications. i.e., a numerical viscosity can be enforced that is in the order of the expected effective viscosity of an eddy viscosity procedure, inspired, e.g., by implicit LES \cite{margolin2006modeling, grinstein2007implicit, adams2009implicit}.


\subsection{Single 3D Flow Rotor in Idealized Conditions with Two End-Plates}
\label{subsec:single_rotor}
This section increases the complexity by conducting a three-dimensional investigation on a full-scale Flettner rotor in idealized wind tunnel conditions. The effort's primary aim refers to the prediction quality of three-dimensional flow effects.

\begin{figure}[!htb]
\begin{minipage}[c][6cm][t]{1.0\textwidth}
	\vspace*{\fill}
	\centering
	\subfigure[]{
	\includegraphics[width=0.475\textwidth]{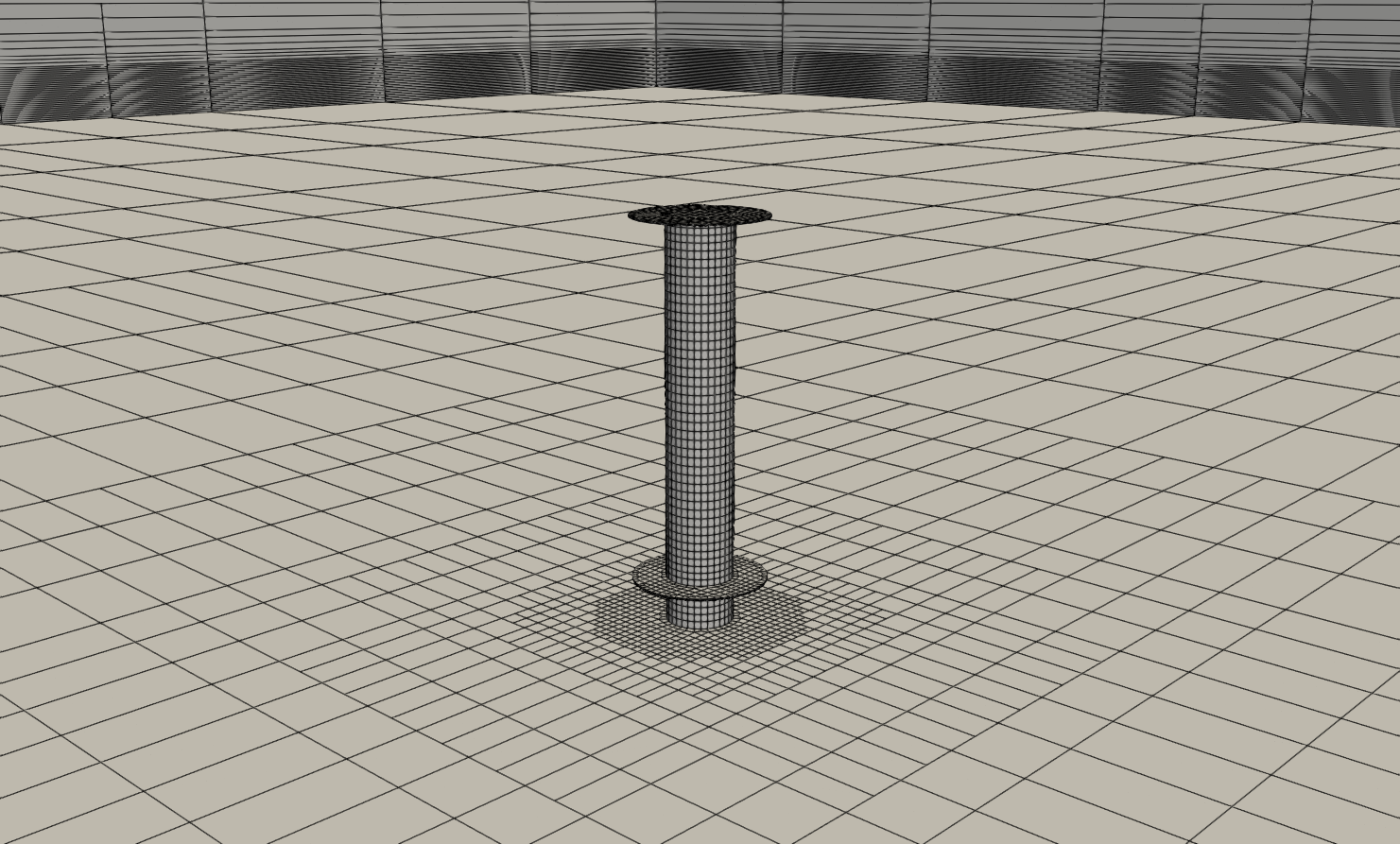}
	}
	\subfigure[]{
	\includegraphics[width=0.475\textwidth]{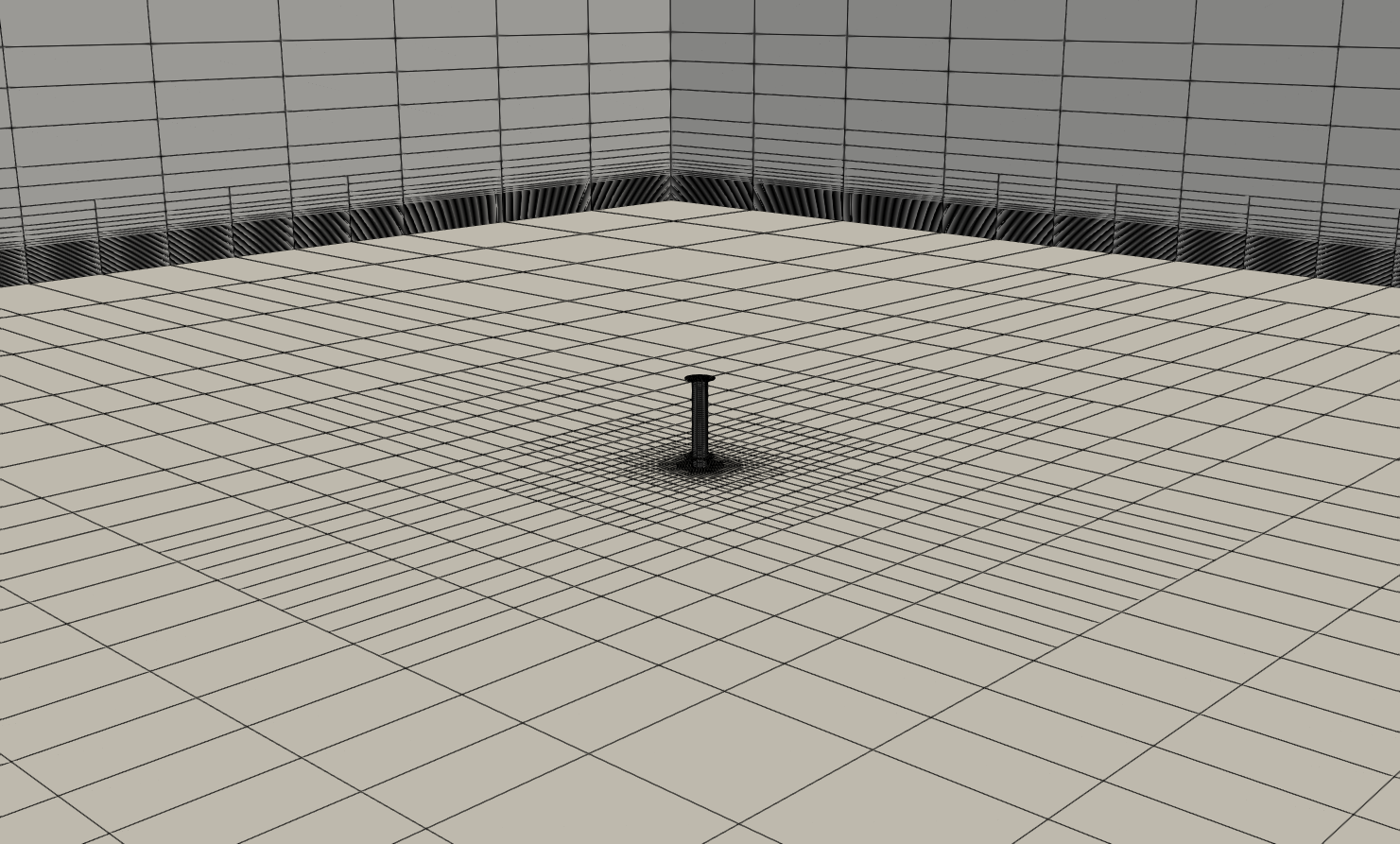}
	}
\end{minipage}
\caption{Single rotor flow: Perspective view of a medium-sized numerical (a) near and (b) far-field grid.}
\label{fig:single_rotor_grid_perspectives}
\end{figure}

The considered rotor features an aspect ratio of $H/D = 6$ and is equipped with two identical upper and lower end plates, whereby the end plate to rotor diameter ratio is $D_\mathrm{endplate}/D_\mathrm{rotor} = 2$. The lower end plate is placed at $H/9$, and both end plates have a height of $H_\mathrm{endplate} / H_\mathrm{rotor} = 0.005$. The far-field boundaries are placed one hundred rotor diameters away from the rotor to minimize the blockage effect. The incoming wind features an atmospheric profile with $h^\mathrm{wind} / H_\mathrm{rotor} = 1/2$ and $\mathrm{mExp} = \SI{0.085}{}$, cf. Alg. \ref{alg:initial_flow_field}. An exemplary impression of a utilized numerical grid of medium resolution with approx. \SI{30000}{} control volumes is provided in Fig. \ref{fig:single_rotor_grid_perspectives} for (a) the near field and (b) the far field. The wind profile is sampled with 20 control volumes.

\begin{figure}[!htb]
\begin{minipage}[c][6cm][t]{1.0\textwidth}
	\vspace*{\fill}
	\centering
	\subfigure[]{
	\includegraphics[width=0.475\textwidth]{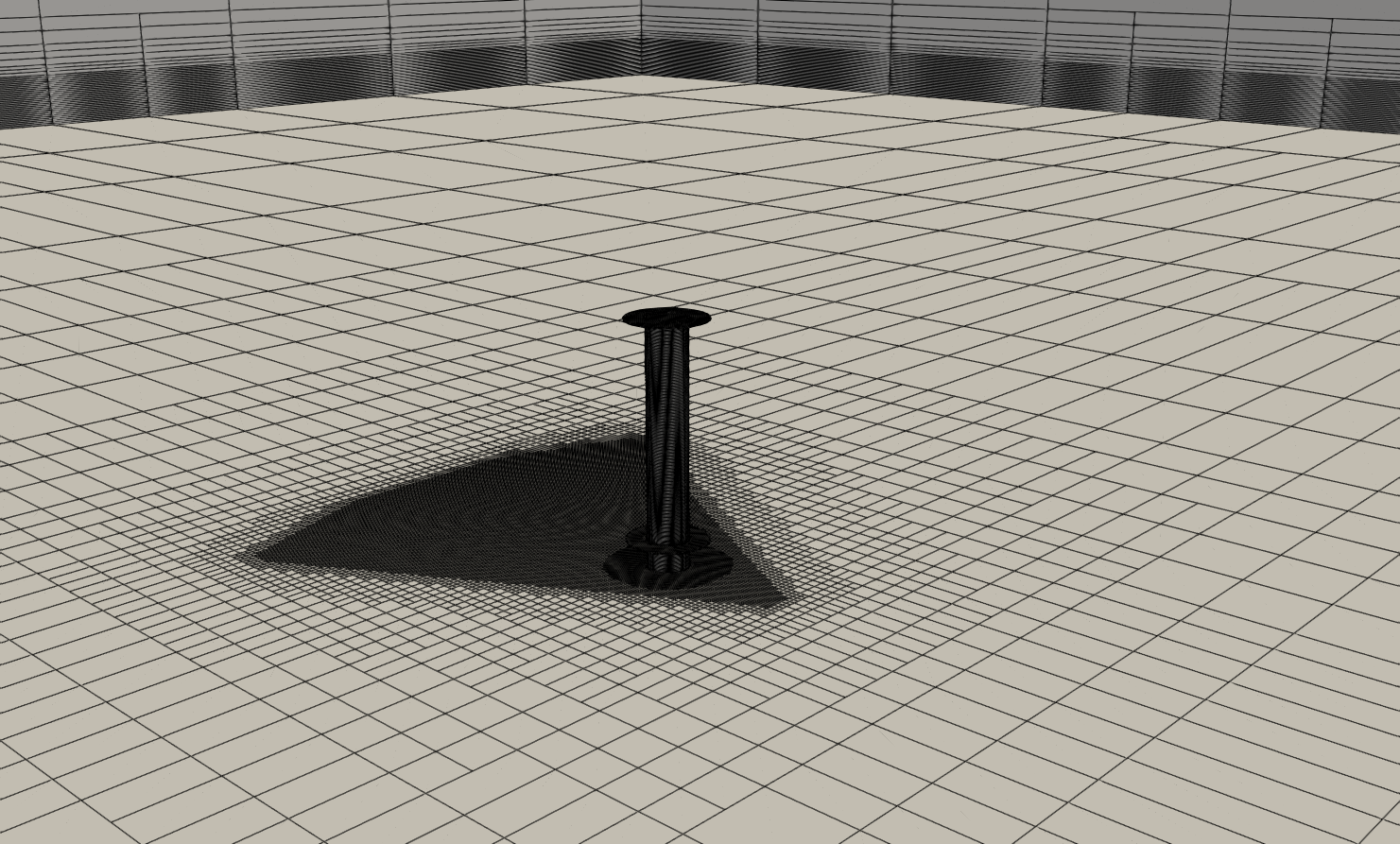}
	}
	\subfigure[]{
	\includegraphics[width=0.425\textwidth]{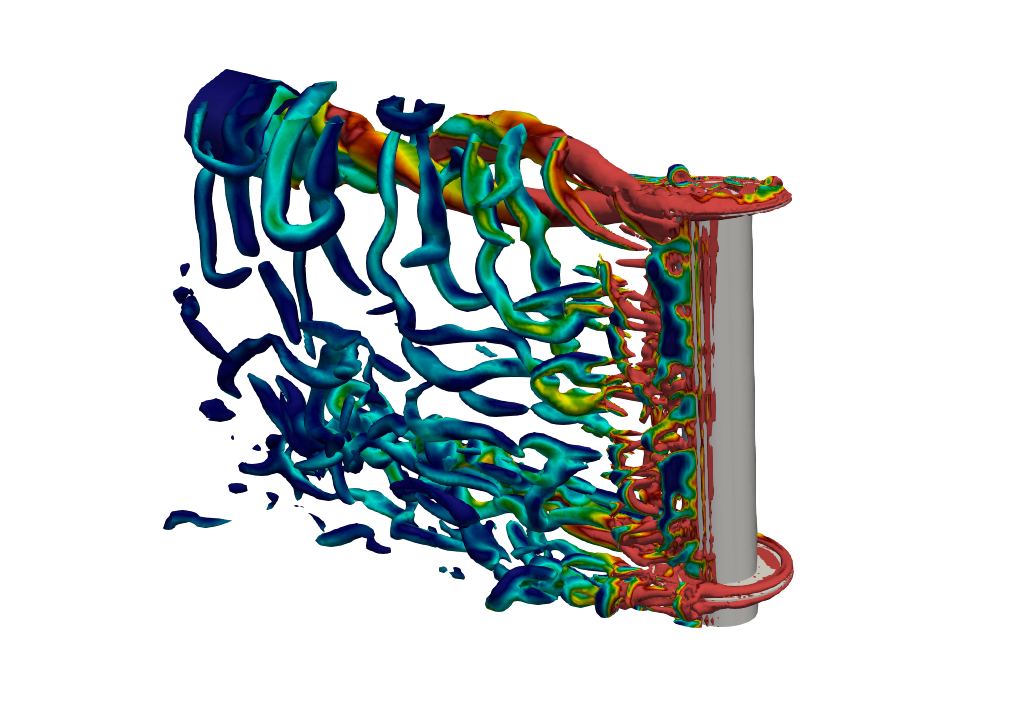}
	}
\end{minipage}
\caption{Single rotor flow: Perspective view of the near-field grid and (b) instantaneous vortex structures colored with the vorticity magnitude of a hybrid averaged/filtered RANS/LES reference simulation at $\mathrm{Re}_\mathrm{H} = 1.2 \cdot 10^7$ or $\mathrm{Re}_\mathrm{D} = 2 \cdot 10^6$ at $k=3.0$.}
\label{fig:single_rotor_DES}
\end{figure}

High-fidelity results from a hybrid averaged/filtered RANS/LES analysis at $\mathrm{Re_D} = 2 \cdot 10^6$ or $\mathrm{Re_H} = 1.2 \cdot 10^7$ serve as benchmark data. An impression of the corresponding numerical grid is shown in Fig. \ref{fig:single_rotor_DES} (a), where the near rotor and the wake region are considerably more refined compared to the Eulerian grid. The grid consists of approx. 11 million control volumes. The figure on the right provides an instantaneous impression of the scale-resolving simulation at $k=3$ based on $Q=100$ iso-lines of the Q-vortex-criterion, where the local vorticity colors the iso-lines. In addition to the smaller vortices in the rotor wake, the tip vortex at the rotor end is particularly recognizable.

In the following, the spinning-ratio spectrum for $0 \leq k \leq 6$ is considered in steps of $\Delta k = 0.5$, and three different strategies for approximating the convective momentum transport are investigated. In addition to the pure UDS method ($\beta=1$), the $\kappa$-scheme for QUICK ($\kappa = 0.5$) and CDS ($\kappa = 1.0$) is assessed. Three numerical grids with different resolutions are considered: While grid 1 discretizes the rotor with about 25 discrete elements along the the rotor diameter and the entire field with about \SI{20000}{} control volumes, grid 2 increases this to 50 as well as \SI{40000}{} and grid three to 100 as well as \SI{80000}{} cells, respectively. Hence, a total of $13$ (spinning ratios) $ \times 3$ (schemes) $ \times 3$ (grids) $ = 117$ Eulerian and 13 scale-resolving simulations are carried out. The Eulerian simulations adjust the rotor circulation every 100 time steps. The time step is set for the Eulerian and scale-resolving simulations so that 100 time steps sample a fluid particle's equivalent rotor passage. Both methods simulate 200 equivalent rotor flows, in which the Eulerian simulations seek pseudo-steady results and thus perform only three outer iterations per time step, cf. results from Sec. \ref{subsec:cirular_cylinder}, e.g., Fig. \ref{fig:rotating_cylinder_vary_nIter_lift}. The scale-resolving simulations have a higher convergence rigor and require a relative residual decrease per equation of at least two orders of magnitude per time step, which, for the chosen relaxation parameter, leads to an increased numerical effort of approx. 10 outer iterations per time step. Local and integral flow data are averaged over the last 20 equivalent rotor flows.

\begin{figure}[!htb]
\centering
\iftoggle{tikzExternal}{
\input{./tikz/02__single_rotor/grid_study_cd_3D.tikz}}{
\includegraphics{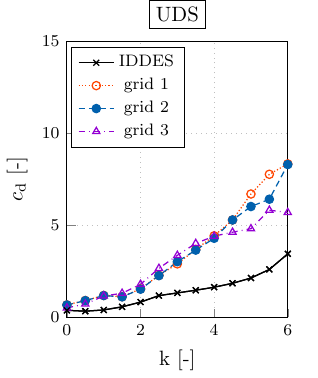}
\includegraphics{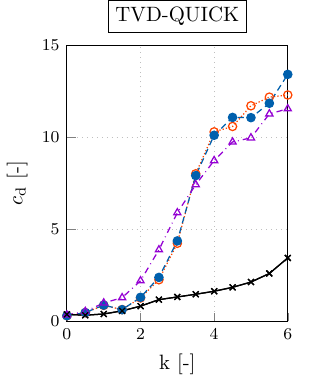}
\includegraphics{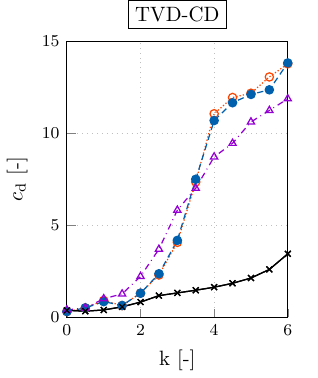}
}
\caption{Single rotor flow: Drag coefficient over the spinning ratio predicted by Eulerian tests on four different grids for a UDS (left), QUICK (center), and CDS (right) approach. In addition, the same viscous reference result is shown.}
\label{fig:single_rotor_cd_3D}
\end{figure}

Figure \ref{fig:single_rotor_cd_3D} presents the predicted drag coefficients over the spinning ratio for all scrutinized grids, distinguishing between the UDS method (left, $\beta = 1$), the QUICK (center, $\kappa = 0.5$), and the CDS method (right, $\kappa = 1.0$) for approximating the convective momentum transport. 
Generally, an increase in the drag coefficient can be observed in all cases for rising spinning ratios. However, all Eulerian results significantly overshoot the viscous reference results. The drag coefficient for the finest considered grid 3 tends to be the smallest, especially for high spinning ratios. Additionally, the drag results of the UDS studies appear to be lower than those of the two $\kappa$-scheme-based results.

\begin{figure}[!htb]
\centering
\iftoggle{tikzExternal}{
\input{./tikz/02__single_rotor/grid_study_cl_3D.tikz}}{
\includegraphics{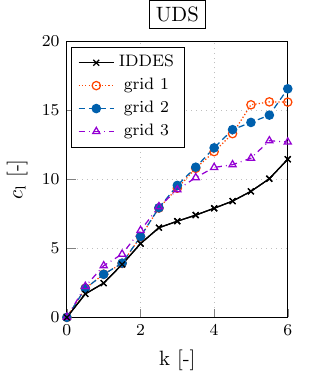}
\includegraphics{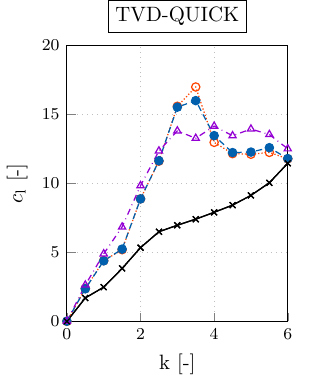}
\includegraphics{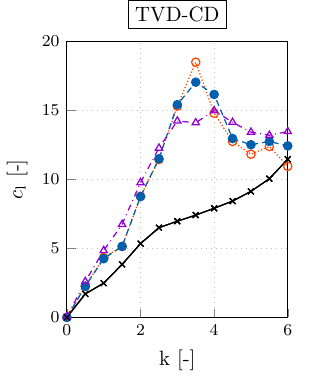}
}
\caption{Single rotor flow: Lift coefficient over the spinning ratio predicted by Eulerian tests on four different grids for a UDS (left), QUICK (center), and CDS (right) approach. In addition, the same viscous reference result is shown.}
\label{fig:single_rotor_cl_3D}
\end{figure}

The predictions of the associated lift coefficients are shown analogously in Fig. \ref{fig:single_rotor_cl_3D}. In line with the drag results, the Eulerian investigations' lift predictions are likewise significantly overestimated compared to the viscous reference results, especially for high spinning ratios. Predictions based on the UDS method increase almost linearly for increasing spinning ratios and are close to the viscous reference results up to approximately $k=2$. Furthermore, in the UDS case, the finest grid provides larger [smaller] lift coefficients for smaller [larger] spinning ratios. The two further investigations based on the $\kappa$-scheme show an almost linear increase with, compared to UDS, an increased slope up to approx. $k=3.5$, which abruptly ends and starts forming a plateau. For large spinning ratios around $k=6$, inviscid predicted lift coefficients approach the viscous reference results. In addition, the predictions of the $\kappa$-scheme-based investigations in the plateau region for $4 \leq k \leq 6$ are of the same order of magnitude as the UDS-based predictions, which indicates an increased switching from QUICK/CDS to UDS, i.e., from high-order to low-order convection. 

\begin{figure}[!htb]
\centering
\iftoggle{tikzExternal}{
\input{./tikz/02__single_rotor/grid_study_cd_2D.tikz}}{
\includegraphics{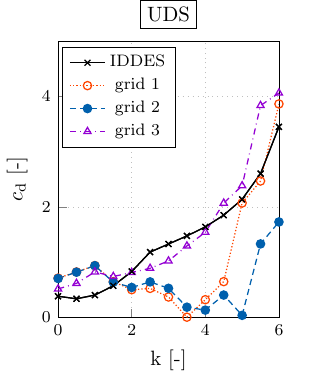}
\includegraphics{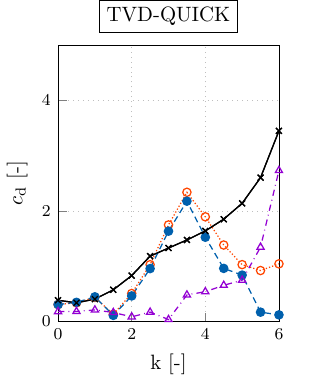}
\includegraphics{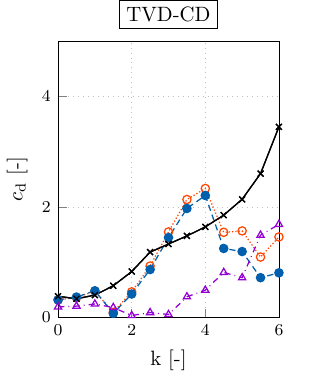}
}
\caption{Single rotor flow (pseudo 2D): Drag coefficient over the spinning ratio predicted by Eulerian tests on four different grids for a UDS (left), QUICK (center), and CDS (right) approach, skipping the momentum balance in the rotor axis direction. In addition, the same 3D viscous reference result is shown.}
\label{fig:single_rotor_cd_2D}
\end{figure}

\begin{figure}[!htb]
\centering
\iftoggle{tikzExternal}{
\input{./tikz/02__single_rotor/grid_study_cl_2D.tikz}}{
\includegraphics{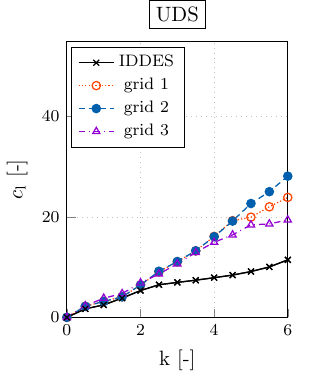}
\includegraphics{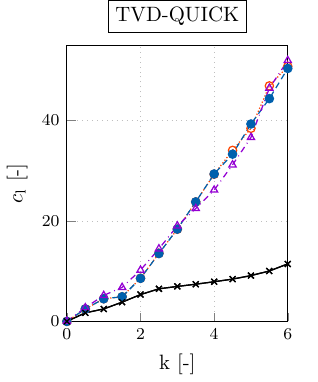}
\includegraphics{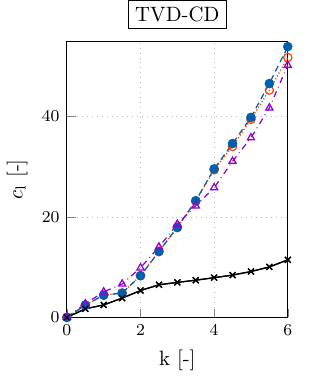}
}
\caption{Single rotor flow (pseudo 2D): Lift coefficient over the spinning ratio predicted by Eulerian tests on four different grids for a UDS (left), QUICK (center), and CDS (right) approach, skipping the momentum balance in the rotor axis direction. In addition, the same 3D viscous reference result is shown.}
\label{fig:single_rotor_cl_2D}
\end{figure}

The fundamentally low parasitic drag for $k=0$, coupled with the significantly increased drag coefficients for $k > 0$ as well as the increased lift coefficient, indicates a dominant influence of induced drag components. Based on considerations from lifting-line theory, this subsection's remainder will examine the quadratic relationship between the lift and induced drag coefficients, i.e., $c_\mathrm{d,i} = \propto c_\mathrm{l}^2$. For this purpose, all Eulerian simulations are repeated, but the approximation of the momentum in the rotor axis direction ($x_3$ direction) is suppressed, resembling a pseudo 2D or 2.5\,D investigation, that should suppress the generation of 3D tip vortices and drive the simulation results closer to flow potential-based results. Analogous to Figs. \ref{fig:single_rotor_cd_3D}-\ref{fig:single_rotor_cl_3D}, Figs. \ref{fig:single_rotor_cd_2D} and \ref{fig:single_rotor_cl_2D} provide the drag and lift coefficients over the considered spinning ratios for the pseudo 2D studies, distinguishing from left to right between UDS, QUICK, and CDS-based investigations. While the drag coefficients are significantly reduced towards the order of magnitude of the viscous reference studies, the lift coefficients are particularly increased and rise almost constantly when increasing the spinning ratio, especially for the $\kappa$-scheme-based investigations. In comparison, the UDS-based methods yield reduced lift but increased drag coefficients. Overall, the pseudo-2D results are similar to the results of the circular cylinder study from the previous Sec. \ref{subsec:cirular_cylinder}.

\begin{figure}[!htb]
\centering
\iftoggle{tikzExternal}{
\input{./tikz/02__single_rotor/induced_resistance.tikz}}{
\includegraphics{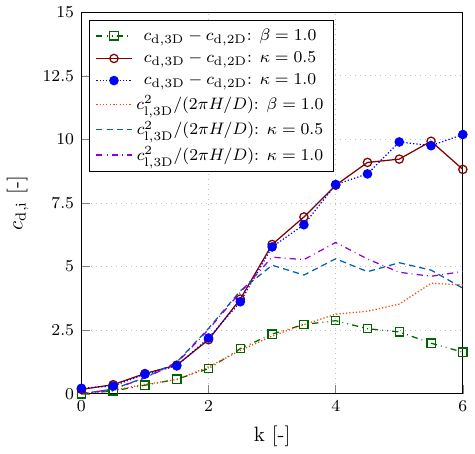}
\includegraphics{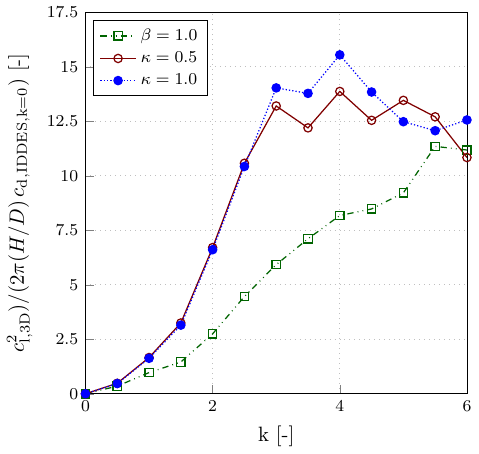}
}
\caption{Single rotor flow: Estimated (left) induced drag for two measures and (right) rotor efficiency based on a parasitic component from an IDDES simulation at rotor rest over the spinning ratio for three discrete approximation strategies on the finest grid.}
\label{fig:single_rotor_efficiency}
\end{figure}

Two results of two induced-drag coefficient $c_\mathrm{d,i}$ estimation approaches are presented in Fig. \ref{fig:single_rotor_efficiency} (left) over the spinning ratio, which (a, with marks) form the difference of the drag coefficients from the 3D (cf. Fig. \ref{fig:single_rotor_cd_3D}) to the 2D (cf. Fig. \ref{fig:single_rotor_cd_2D}) investigations, i.e., $c_\mathrm{d, 3D} - c_\mathrm{d, 2D}$ and (b, without marks) assume a scaled, squared correlation with the lift coefficient. The latter argumentation assumes a vanishing parasitic resistance at, i.e., $c_\mathrm{d,2D} (k=0) = 0$ in line with lifting line arguments. All three investigated convection schemes are considered on the finest numerical grid. Both induced drag measures predict almost identical values up to a spinning ratio of $k=3$ and only start deviating for higher spinning ratios, with the bifurcation in the UDS investigations even being shifted to approx. $k=4$. For high spinning ratios, all induced drag estimates based on the lifting line theory tend towards a value of $c_\mathrm{d,i} \approx 5$. The 3D-2D difference results of the $\kappa$-scheme investigations overshoot this and reach twice as high values around $c_\mathrm{d,i} \approx 10$. In contrast, the values of the UDS investigations drop to $c_\mathrm{d,i} \approx 2$.
Finally, the estimated induced drag coefficients are related to their parasitic companion to estimate a realized rotor efficiency in Fig. \ref{fig:single_rotor_efficiency} (right). The metric is similar to the Oswald factor of classic lifting line assumptions. The parasitic reference resistance is taken from the IDDES investigation on the non-rotating rotor cf. black reference values in Figs. \ref{fig:single_rotor_cd_3D} and \ref{fig:single_rotor_cd_2D} at $k=0$, multiplied by the rotor's aspect ratio times two pi. Both rotor efficiencies of the $\kappa$-scheme-based investigations are again similar and of the same order of magnitude. They reach an almost constant value of $c_\mathrm{l}^2 / 2 \pi (H/D) c_\mathrm{d,IDDES,k=0} \approx 14$ for $k \geq 3$. In contrast, the efficiency of the UDS-based investigations is reduced, and $ c_\mathrm{l}^2 / 2 \pi (H/D) c_\mathrm{d,IDDES,k=0} \approx 11$ is reached for the highest spinning ratios $k \geq 5.5$ considered only. Compared to classic profile flows, the comparatively high rotor efficiency can be attributed to the additional momentum input of the spinning rotor.


\paragraph{A summary in between:} Since the high-order-convection methods based on the $\kappa$-scheme introduce significantly less numerical viscosity into the discrete system, these scenarios provide significantly increased lift coefficients, which are close to the flow-potential-based reference results for academic 2D investigations on a circular cylinder. However, in a three-dimensional context, the presented method can adequately reproduce induced resistances, especially the quadratic relation between induced drag and lift, i.e., $c_\mathrm{d} \propto c_\mathrm{l}^2$, which in turn leads to a solid overestimation of drag coefficients, especially for high spinning ratios. The lift and drag overestimation is reduced for low-order-convection-based approaches, introducing increased numerical viscosity. Influences of the spatial discretization cannot be excluded entirely. However, the influence of the convection scheme seems significantly increased, especially since the primary purpose refers to a rough relative rotor performance prediction.

\subsection{3D Flow Around a General Cargo Ship at Realistic Operation Conditions}
\label{sec:annika_braren}




Next, the complexity of the validation is further increased by examining a real operating vessel equipped with Flettner rotors. 
For the paper's analysis, accurate ship data is used to investigate suitable operating conditions: The vessel is traveling at 6.8 knots with a draft of 6m without trim in a wind of $\SI{8}{m/s}$ (Beaufort 5). The wind profile is approximated using an atmospheric profile with $h^\mathrm{wind} = \SI{20}{m}$ as well as $\mathrm{mExp} = \SI{0.085}{}$ (cf. Alg. \ref{alg:initial_flow_field}), and the True/Apparent Wind Angle (T/AWA) of attack varies in the study.
Based on the operating data, the ship cruises at $\mathrm{Re}_\mathrm{L_{pp}} = V_\mathrm{wind} L_\mathrm{pp} / \nu_\mathrm{air} \approx \SI{1.039E+08}{}$ and the rotor of diameter $D$ operates at $k = \pi \, n \, D / V_\mathrm{wind} \approx 2.4$, where $\nu_\mathrm{air}$, $L_\mathrm{pp}$, and $V$ refer to the dynamic viscosity of air, the ship's length between the perpendiculars and a representative apparent wind magnitude 10m over ground, respectively.

It should be noted that the considered rotor configuration, including a lower endplate, reflects a real-world installation on an existing vessel. While many Flettner rotor systems use the ship deck as a natural endplate, additional lower plates may be present in practical implementations. The present study intentionally retains this configuration to represent such real-world conditions.

\begin{figure}[!htb]
\begin{minipage}[c][11cm][t]{0.3\textwidth}
	\vspace*{\fill}
	\centering
	\subfigure[]{
	\includegraphics[width=0.9\textwidth]{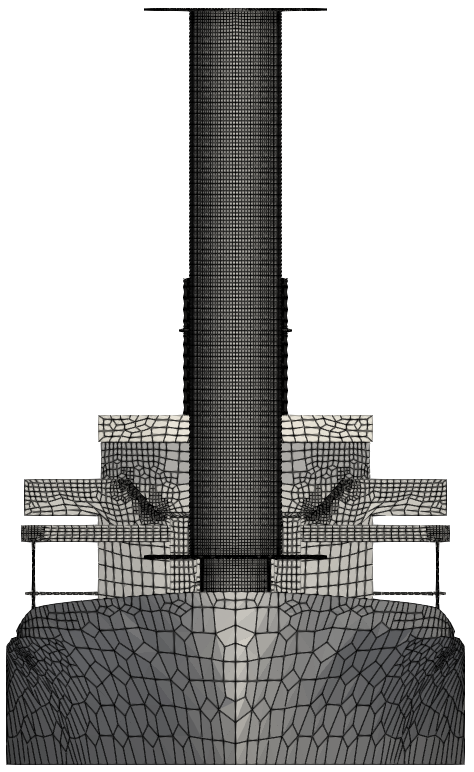}
	}
\end{minipage}
\begin{minipage}[c][11cm][t]{0.7\textwidth}
	\vspace*{\fill}
	\centering
	\subfigure[]{
	\includegraphics[width=1.0\textwidth]{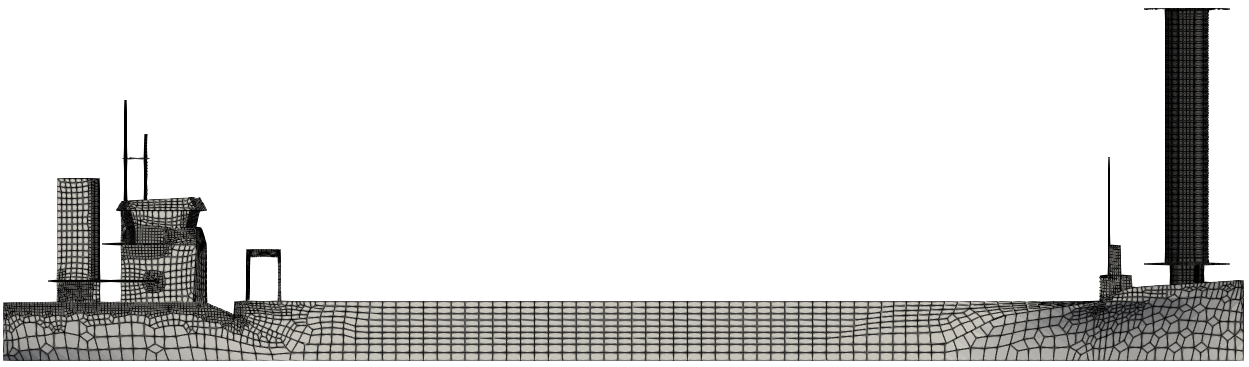}
	}
	\subfigure[]{
	\includegraphics[width=1.0\textwidth]{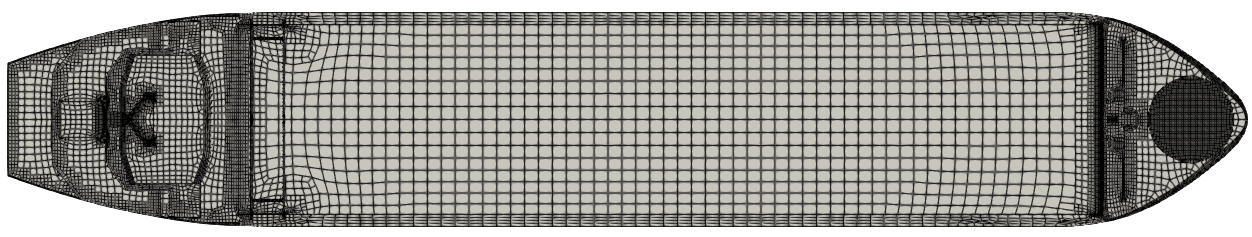}
	}
\end{minipage}
\caption{General cargo ship flow: The utilized surface discretization presented from (a) front, (b) starboard side, and (c) top.}
\label{fig:annika_braren_grid_x1x2x3}
\end{figure}

\begin{figure}[!htb]
\begin{minipage}[c][11cm][t]{1.0\textwidth}
	\vspace*{\fill}
	\centering
	\subfigure[]{
	\includegraphics[width=0.475\textwidth]{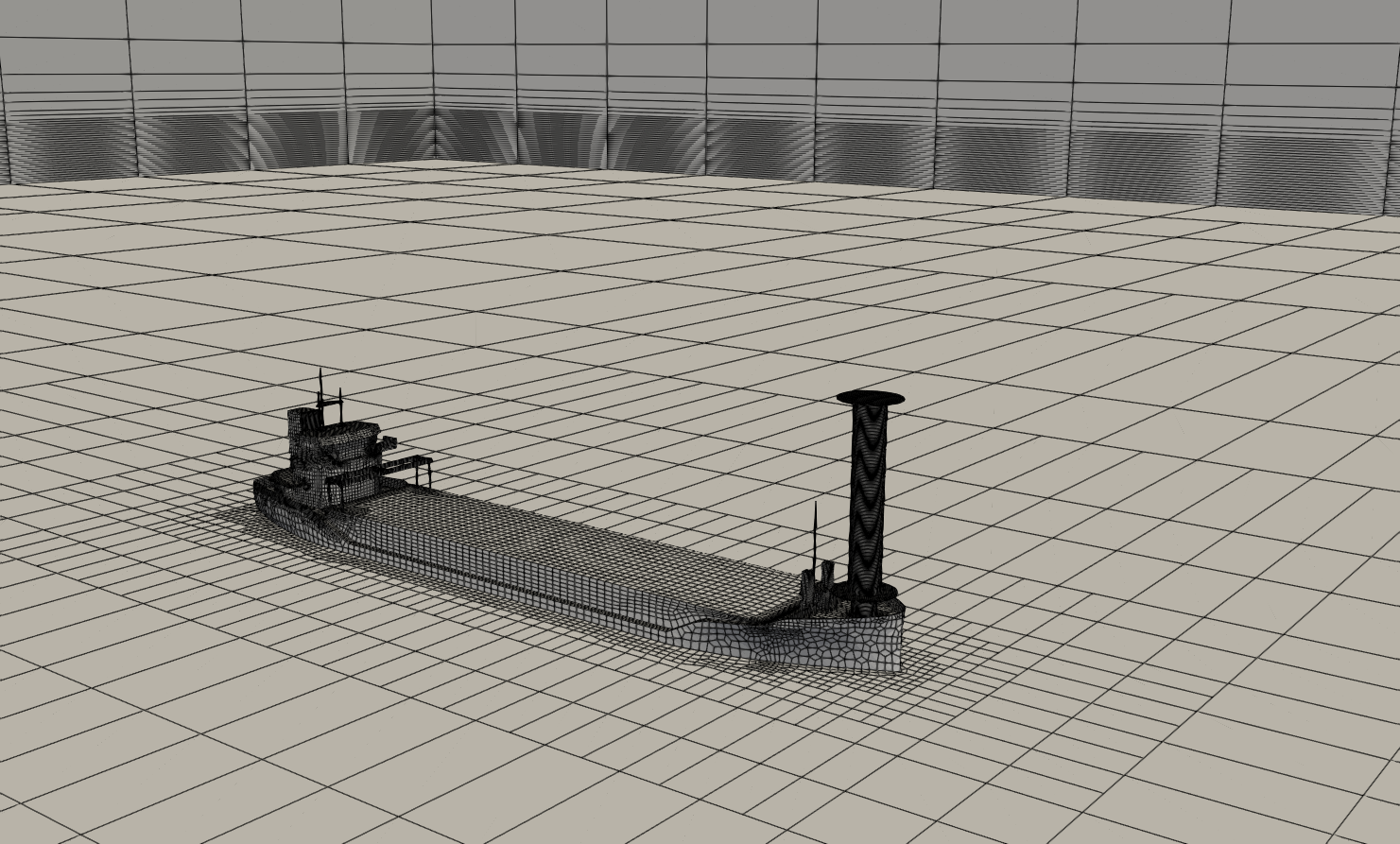}
	}
	\subfigure[]{
	\includegraphics[width=0.475\textwidth]{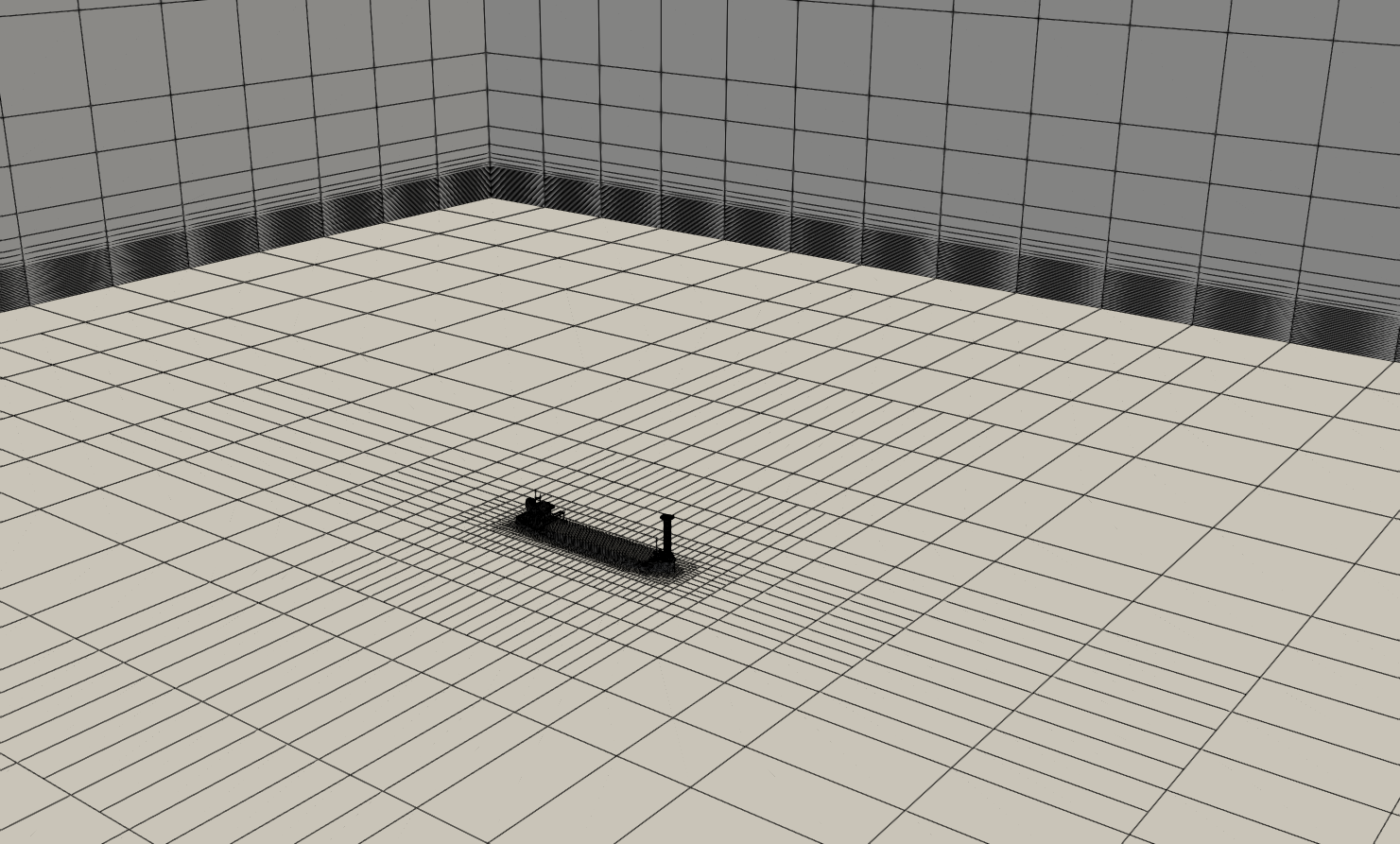}
	}
\end{minipage}
\caption{General cargo ship flow: Perspective view of the numerical (a) near and (b) far-field grid.}
\label{fig:annika_braren_perspectives}
\end{figure}

The investigations are carried out on three unstructured numerical grids of varying fineness along and near the rotor. The ship is discretized in the longitudinal direction with approx. 225 cells, whereby finer geometric details are treated more precisely. In general, the grid is coarsened away from the ship. The three different grids vary only in their resolution of the Flettner rotor, which is discretized with either 25, 50, or 100 cells in the circumferential and normal directions so that the three grids consist of approx. \SI{585000}{}, \SI{695000}{}, and \SI{995000}{} cells. An impression of the surface ship is provided in Fig. \ref{fig:annika_braren_grid_x1x2x3} for (a) the front, (b) the starboard side, and (c) the top view, where the surface discretization of the second grid refinement is depicted. The corresponding discretization of the outer far-field is shown in Fig. \ref{fig:annika_braren_perspectives}, which also indicates the refinement in the atmospheric wind profile region.
The investigated domain extends over a length, width, and height of 11, 11, and 5 ship lengths, respectively, with the ship placed centered at the lower end. The aerodynamic investigations are based on an undeflected free water surface, which is treated as a symmetry boundary condition at the lower end of the domain. A constant pressure is specified at the upper end, and the apparent wind (i.e., superposition of ship and wind speed) is specified as a superposition along all horizontal outer boundaries, cf. Alg. \ref{alg:initial_flow_field}. Initial conditions follow the far field apparent wind boundary conditions. The time step is chosen such that an equivalent flow around the ship is sampled with approximately 150-time steps. Each inviscid simulation is performed for $T= \SI{3000}{}$ time steps. Results are averaged over the last 10 equivalent flows, i.e., \SI{1500}{} time steps.

\begin{figure}[!htb]
\begin{minipage}[c][11cm][t]{1.0\textwidth}
	\vspace*{\fill}
	\centering
	\subfigure[]{
	\includegraphics[width=0.475\textwidth]{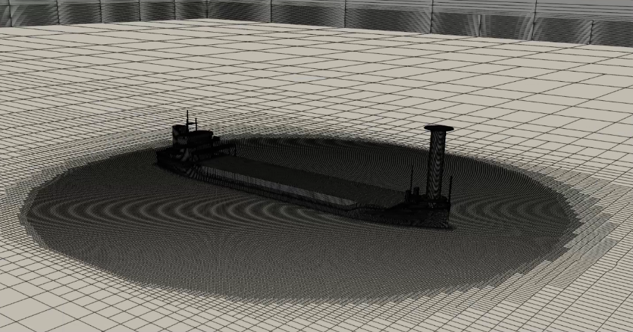}
	}
	\subfigure[]{
	\includegraphics[width=0.475\textwidth]{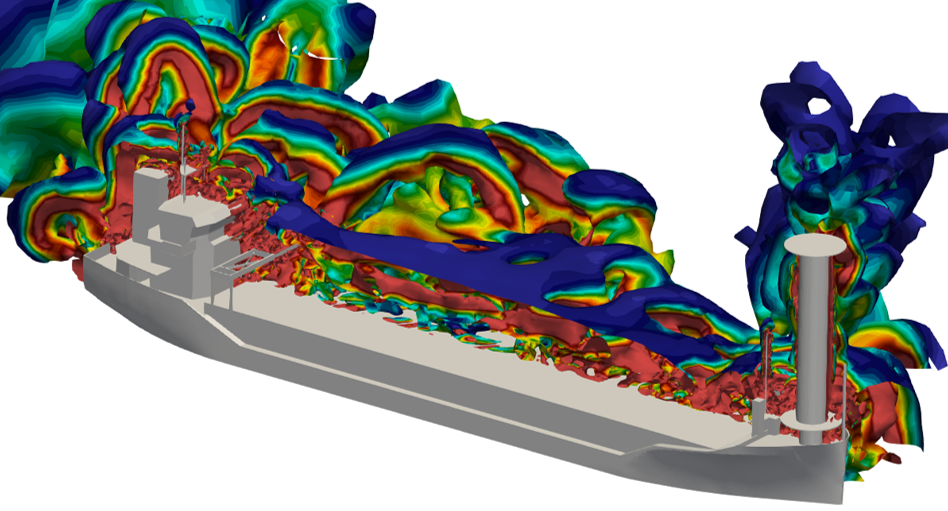}
	}
\end{minipage}
\caption{General cargo ship flow: Perspective view of the near-field grid and (b) instantaneous vortex structures colored with the vorticity magnitude of a hybrid averaged/filtered RANS/LES reference simulation at $\mathrm{Re}_\mathrm{L} = 1 \cdot 10^8$ and $\mathrm{Re}_\mathrm{D} = 3.45 \cdot 10^6$.}
\label{fig:annika_braren_DES}
\end{figure}

To assess the quality of the method proposed in this paper, reference results on the general cargo vessel follow hybrid averaged/filtered RANS/LES simulations. Their spatial and temporal discretization differs significantly from the inviscid investigations. A perspective impression of the utilized surface grid is given in Fig. \ref{fig:annika_braren_DES} (left) and indicates an intensified refinement in regions close to the ship. The refinement sphere around the ship's midpoint is discretized with $\Delta x_1 / L_\mathrm{pp} = \Delta x_2 / L_\mathrm{pp} = \Delta x_3 / L_\mathrm{pp} = 1/225$. In combination with the additional required refinement in the wall-normal direction, which allows the use of wall functions by suitable values of $y^+ \approx 50$, this results in an increased numerical grid with approximately eleven million control volumes. The time step is reduced to comply with an averaged Courant number below one, resulting in an increased number of $T = \SI{90000}{}$ time steps. In each time step, all residuals are reduced by at least two orders of magnitude. Further details, such as an increase of the temporal approximation order by a third time level or the use of 20/80 percent QUICK/CDS ($\kappa = 0.9$ in Eqn. \eqref{equ:kappa_scheme}) can be found in \cite{angerbauer2020hybrid, pache2023datenbasierte}. Therein, the method is validated in particular against experimental wind tunnel tests, which underlines its fundamental suitability for providing high-fidelity reference data for the simulation strategy pursued in this paper. An instantaneous impression of the scale-resolving simulations is shown in Fig. \ref{fig:annika_braren_DES} based on vortex structures colored with the vorticity magnitude.

In addition to the IDDES studies, further Unsteady RANS (URANS) studies in line with the 2003 version of Menter's SST model are also conducted, see \cite{menter2003ten}. Results of both viscous methods largely align, which is why only the IDDES simulations are used as a reference and the URANS simulations only classify the numerical effort at this section's end.

\begin{figure}[!htb]
\centering
\iftoggle{tikzExternal}{
\input{./tikz/03__annika_braren/exemplary_results.tikz}}{
\includegraphics{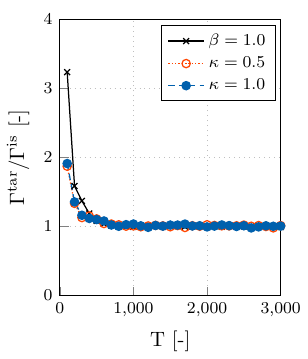}
\includegraphics{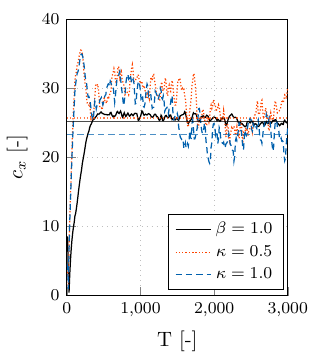}
\includegraphics{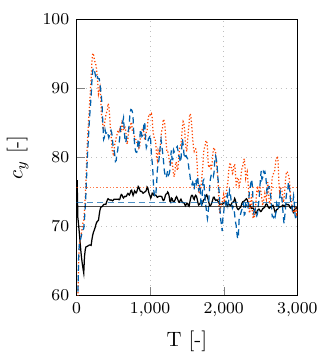}
}
\caption{General cargo ship flow : Exemplary development of (left) the ratio of the target/actual flow circulation, (center) the resistance, and (right) the side force coefficient over the simulated time steps for three different treatments of the convective momentum transport at a true wind angle of $90^\circ$. Horizontal lines indicate the average resistance values obtained from averaging over the last 1050 time steps.}
\label{fig:annika_braren_exemplary_results}
\end{figure}

Based on the two-dimensional cylinder study from Sec. \ref{subsec:cirular_cylinder}, an impression of the rudimentary behavior of the dynamic circulation adjustment procedure is first given. For this purpose, Fig. \ref{fig:annika_braren_exemplary_results} shows the development of the ratio of target to the actual rotor circulation from Alg. \ref{alg:body_force_update} (left), the longitudinal (center), and the lateral (right) wind coefficient for a TWA of 90 degrees approaching from starboard, evaluated on the second grid level. Three different convection schemes are considered: a UDS (solid black), a TVD-QUICK (orange dotted), and a TVD-CDS (blue dashed) procedure. Again, the volume force is initially undersized and needs to be increased. The discrete adjustment scenarios can be identified in the diagram by the discrete markers every $T = 100$ time steps and again converge quickly to the desired unit value within approximately 10 adjustments. The resulting forces acting on the superstructure are subject to fluctuations, even for nearly constant circulation ratios and, thus, volume forces. The fluctuations are significantly reduced for the UDS simulations. However, all three convection schemes result in similar forces averaged over the last 1500 time steps, which are indicated as horizontal lines. Predictions between UDS and TVD-QUICK are close to each other for the longitudinal forces; in the case of the lateral forces, a similarity between UDS and TVD-CDS occurs, which, e.g., underlines the increased lift predictions for TVD-QUICK compared to UDS or TVD-CDS, cf. Fig. \ref{fig:rotating_cylinder_vary_nIter_lift}.

\begin{figure}[!htb]
\centering
\iftoggle{tikzExternal}{
\input{./tikz/03__annika_braren/cx_cy_TWA_grids.tikz}}{
\includegraphics{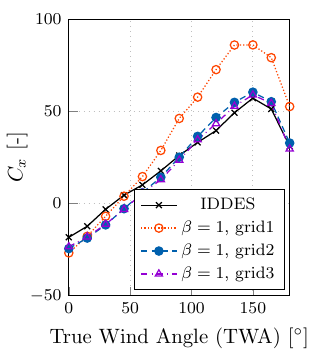}
\includegraphics{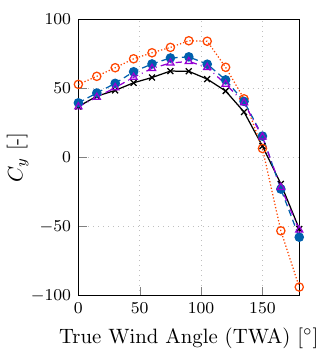}
\includegraphics{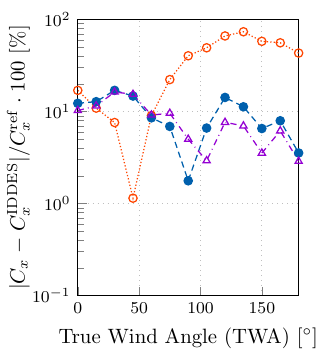}
}
\caption{General cargo ship flow: Longitudinal (left) and lateral (center) wind coefficient, as well as the relative longitudinal wind coefficient error (right) over the True Wind Angle for the reference viscous IDDES, as well as three Eulerian studies with differently sized numerical grids.}
\label{fig:annika_braren_cx_cy_TWA_grids}
\end{figure}

The exemplary study is continued for half the wind polar diagram, i.e., the wind spectrum $0^\circ \leq \mathrm{TWA} \leq 180^\circ$, sampled in 15-degree steps, resulting in 13 simulations to consider the wind spectrum. First, the three introduced numerical grids with different rotor discretization are considered using the flux-blending method with UDS ($\beta = 1.0$) and the hybrid RANS/LES method, resulting in 13x4=52 simulations. Figure \ref{fig:annika_braren_cx_cy_TWA_grids} displays the longitudinal (left, $C_x = 2 f_1 / (\rho V^2 A_\mathrm{f})$) and lateral (center, $C_y = 2 f_1 / (\rho V^2 A_\mathrm{l})$) wind coefficients over the TWA, where $A_\mathrm{f}$ and $A_\mathrm{l}$ refer to the projected frontal and lateral reference area, respectively. Furthermore, the relative error in the longitudinal forces between the inviscid and the IDDES analyses is shown on the right. Generally, the Eulerian and IDDES predictions have a solid qualitative agreement. Only the results based on the coarsest grid significantly underperform, leading to errors on $\mathcal{O}(10^2\%)$. The differences are particularly pronounced in $90^\circ \leq \mathrm{TWA} \leq 180$ for the longitudinal force and $0^\circ \leq \mathrm{TWA} \leq 120^\circ$ as well as $160^\circ \leq \mathrm{TWA} \leq 180^\circ$ for the lateral companion, i.e., precisely the ranges where the rotor generates a force in the respective direction.
%

\begin{figure}[!htb]
\centering
\iftoggle{tikzExternal}{
\input{./tikz/03__annika_braren/cx_cy_AWA_grids.tikz}}{
\includegraphics{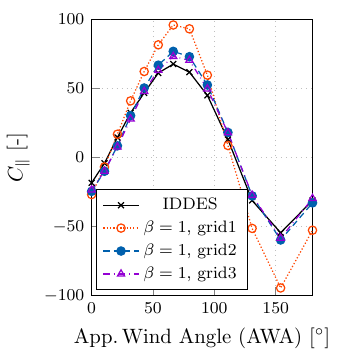}
\includegraphics{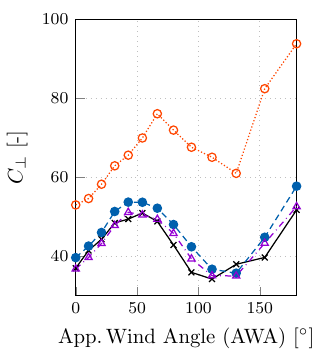}
\includegraphics{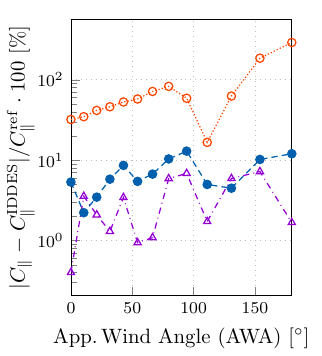}
}
\caption{General cargo ship flow: Wind parallel drag (left) and wind orthogonal side (center) force coefficient, as well as the relative wind coefficient error (right) over the Apparent Wind Angle for the reference viscous IDDES, as well as three Eulerian studies with differently sized numerical grids.}
\label{fig:annika_braren_cx_cy_AWA_grids}
\end{figure}

The deficiency of the coarse-grid procedure in predicting lift as a function of the TWA is already noticeable in Fig. \ref{fig:annika_braren_cx_cy_TWA_grids}. However, it will be further confirmed below by employing the apparent wind scenario. For this purpose, the wind force coefficients are projected into the apparent wind direction, and the resulting wind parallel and wind orthogonal components are displayed in Fig. \ref{fig:annika_braren_cx_cy_AWA_grids} on the left and in the center, respectively. The predictions of the cross-wind forces of the coarsest grid deviate significantly from the forces on the finer grids and the IDDES reference results, underscoring the importance of sufficient rotor refinement for adequately predicting lift effects. In addition, the relative errors of the wind-orthogonal forces between the inviscid and inviscid investigations are shown in the right part of the figure. While the two finer grids again lead to errors of $\mathcal{O}(10^1\%)$, the coarsest grid errors tend to be one magnitude higher.

\begin{figure}[!htb]
\centering
\iftoggle{tikzExternal}{
\input{./tikz/03__annika_braren/cx_cy_TWA_grid2.tikz}}{
\includegraphics{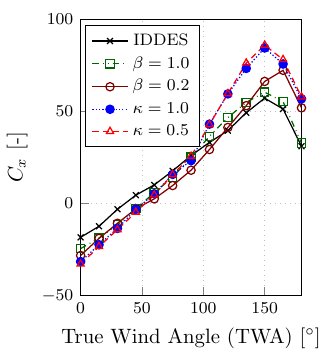}
\includegraphics{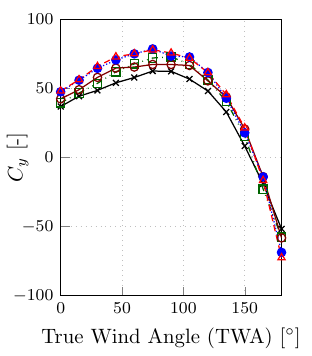}
\includegraphics{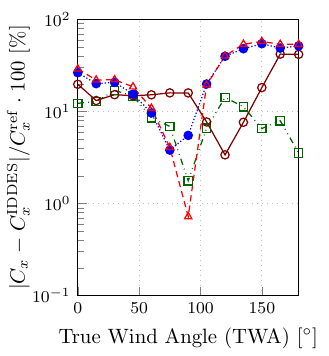}
}
\caption{General cargo ship flow: Longitudinal (left) and lateral (center) wind coefficient, as well as the relative wind coefficient error (right) over the True Wind Angle for the reference viscous IDDES, as well as four Eulerian studies with different approximations of the convective momentum transport.}
\label{fig:annika_braren_cx_cy_TWA_grid2}
\end{figure}

\begin{figure}[!htb]
\centering
\iftoggle{tikzExternal}{
\input{./tikz/03__annika_braren/cx_cy_AWA_grid2.tikz}}{
\includegraphics{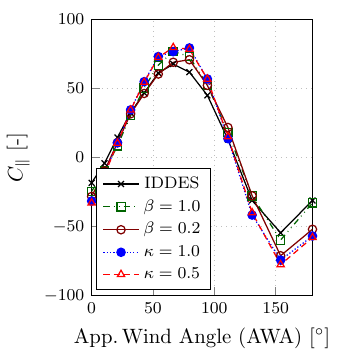}
\includegraphics{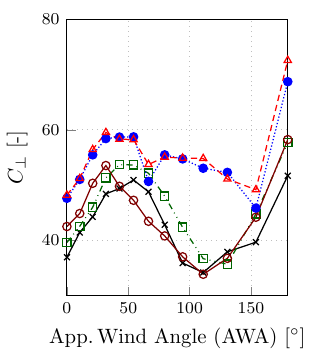}
\includegraphics{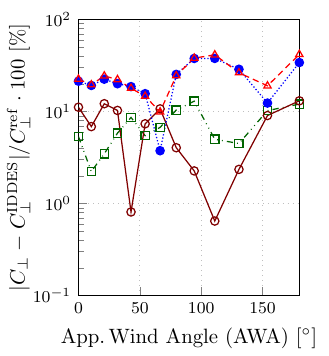}
}
\caption{General cargo ship flow: Wind parallel drag (left) and wind orthogonal side (center) force coefficient, as well as the relative wind coefficient error (right) over the Apparent Wind Angle for the reference viscous IDDES, as well as three Eulerian studies with different approximations of the convective momentum transport.}
\label{fig:annika_braren_cx_cy_AWA_grid2}
\end{figure}

The following investigations are conducted exclusively on the finest grid. In addition to the three considered convection schemes already outlined in Fig. \ref{fig:annika_braren_exemplary_results}, a flux-blending method with 20/80 percent UD/CD is added to the analysis. The four different numerical strategies are re-investigated for all 13 wind scenarios, resulting in 13x3=39 additional Eulerian simulations. The results are presented in Figs. \ref{fig:annika_braren_cx_cy_TWA_grid2} and \ref{fig:annika_braren_cx_cy_AWA_grid2}, and the presentation aligns with the true and apparent wind scenarios in Figures \ref{fig:annika_braren_cx_cy_TWA_grids}-\ref{fig:annika_braren_cx_cy_AWA_grids}. The longitudinal or drag force components are again shown on the left, the lateral or lift force components in the middle, and error measures on the right. It can be noted that the prediction differences get smaller than those in the previous coarse-grid UD studies. Although all lateral forces in Fig. 2 broadly overestimate the IDDES reference results, they generally follow the qualitative curve. A similar picture appears for the longitudinal forces, which only deviate qualitatively from the reference results for larger TWA and the TVD method, resulting in errors in the up to ten percent for the medium grid. The apparent wind scenario in Fig. \ref{fig:annika_braren_cx_cy_AWA_grid2} shows that the differences again originate from lift effects, which are still close to, but slightly out of phase with, the reference results for the flux-blending methods, leading to significantly inflated lift predictions. 




Finally, a rough comparison of the numerical effort is conducted. For this purpose, Tab. \ref{tab:annika_braren_numerical_effort} lists basic discretization as well as averaged, measured simulation effort of all three simulation methods used. In addition to the spatial ($\#$nCells) and temporal ($\#$nTimeSteps) discretization count, the table also provides the degree of spatial parallelization in line with a domain decomposition strategy (\cite{kuhl2024incremental}) distributed across processes ($\#$nCPU) and the required wall clock time $\bar{t}^\mathrm{sim}$. The latter values correspond to averaged values from selected previous studies. A combination of process count and averaged simulation wall clock time allows for the estimation of the numerical effort in terms of CPU core hours, i.e., $\mathrm{CPUh} = \mathrm{nCPU} (\bar{t}^\mathrm{sim} / 3600)$.
Although the URANS method is investigated with the same spatial but approximately ten times coarser temporal discretization, the numerical effort is only approximately one-third as high as the IDDES effort, attributable, e.g., to increased iterative effort per time step due to the increased modeling effort.
\begin{table}[!ht]
\caption{General cargo ship flow: Discretization and simulation data of all three utilized simulation methods as well as resulting, required numerical effort.}
\begin{center}
\begin{tabular}{lccccc}
\toprule
approach    & nCells [-]& nTimeSteps [-]& nCPU [-]  & $\bar{t}^\mathrm{sim}$ [s]& CPUh [-]  \\ \midrule
IDDES       & \SI{11144780}{}  & \SI{90000}{}         & 256       & \SI{0.503E+06}{}          & \SI{35769}{}     \\
URANS       & \SI{11144780}{}  & \SI{9000}{}          & 256       & \SI{0.191E+06}{}          & \SI{13596}{}     \\
Eulerian    & \SI{994746}{}    & \SI{3000}{}          & 96        & \SI{0.871E+03}{}          & \SI{23}{}        \\
\bottomrule
\end{tabular}
\end{center}
\label{tab:annika_braren_numerical_effort}
\end{table}
The numerical effort of the Eulerian methods is reduced by around three orders compared to that of the viscous methods. 

The comparison with viscous reference simulations reveals noticeable deviations in the predicted force magnitudes, particularly at higher spinning ratios and depending on the chosen convection scheme. These discrepancies are primarily attributed to the inviscid formulation and the associated numerical modeling assumptions.
Despite quantitative deviations in absolute force predictions, the method preserves consistent qualitative trends across the investigated parameter space. This makes it suitable for comparative assessments and preliminary ranking of design configurations in early-stage studies, where relative performance differences are of primary interest.


\subsection{Two Interacting Rotors}
\begin{figure}[!ht]
    \centering
    \iftoggle{tikzExternal}{
    \input{./tikz/04__two_rotors/convention_scetch.tikz}}{
    \includegraphics{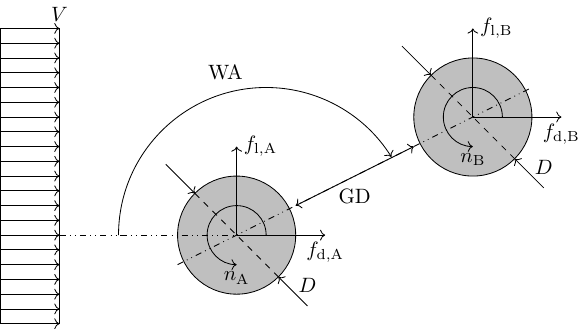}}
    \caption{Two interacting rotors case: Schematic representation of the overall setup in addition to the single cylinder setup from Fig. \ref{fig:rotating_cylinder_2D}.}
    \label{fig:interacting_rotors_scetch}
\end{figure}

In addition to the investigations of the previous section, a typical design space exploration also features the analysis of multi-rotor systems, which leads to further rotor-rotor interactions. This final validation study examines the method's fundamental ability to appropriately represent these rotor-rotor interaction effects. The analysis is inspired by recent wind tunnel experiments (see \cite{bordogna2020aerodynamics}), and a 2D illustration of the study's relative rotor positioning is provided in Fig. \ref{fig:interacting_rotors_scetch}. In addition to the academic studies from Sec. \ref{subsec:cirular_cylinder}, a second rotor (denoted by rotor B) is placed with different Gap Distances (GD) relative to the initial rotor (denoted by rotor A) under varying Wind Angles (WA), with rotor B moving forward on the starboard side of rotor A, i.e., $0^\circ \leq \mathrm{WA} \leq 180^\circ$. For an angle of $\mathrm{WA}=180^\circ$, rotor A is the front rotor seen from the wind, and rotor B is the rear rotor in rotor A's wake. For $\mathrm{WA}=0^\circ$, the situation is reversed. In the case of $\mathrm{WA}=90^\circ$, both rotors are next to each other and experience the bulk flow. Both rotors are identical, have a diameter of $D$, an aspect ratio of $H/D = 6$, and are equipped with an upper-end plate twice the rotor diameter $D_\mathrm{EP}$, i.e., $D_\mathrm{EP} / D = 2$. The wind is identical for all tests and has no height profile but a block profile with a wind speed of $V$. While both rotors have the same direction of rotation, their respective rotational speeds vary.
%
%
%
%
%
%
%

The following parameter variation is carried out: Three different Gap Distances $\mathrm{GD} = [2, 6.5, 14] D$ are examined, whereby the wind angle spectrum is scanned in 15-degree steps, i.e., $\mathrm{WA} = [0, 15, 30, ..., 180]^\circ$, resulting in a total of $3 \times 13 = 39$ rotor-rotor configurations to be investigated. The aerodynamic scenario consists of a constant rotational speed for rotor A and three different rotational speeds of rotor B, i.e., $k_\mathrm{A} = 1$ and $k_\mathrm{B} = [0, 1.5, 3]$. Hence, three simulations are performed for each configuration, resulting in $3 \times 39 = 117$ simulations for an entiree study. In addition to the Eulerian flow approach, benchmark results are generated based on URANS studies, resulting in two complete studies to be conducted with an overall amount of $2 \times 117 = 234$ simulations plus two additional stand-alone considerations.

\begin{figure}[!htb]
    \vspace*{\fill}
    \centering
    \subfigure[]{
    \includegraphics[width=0.315\textwidth]{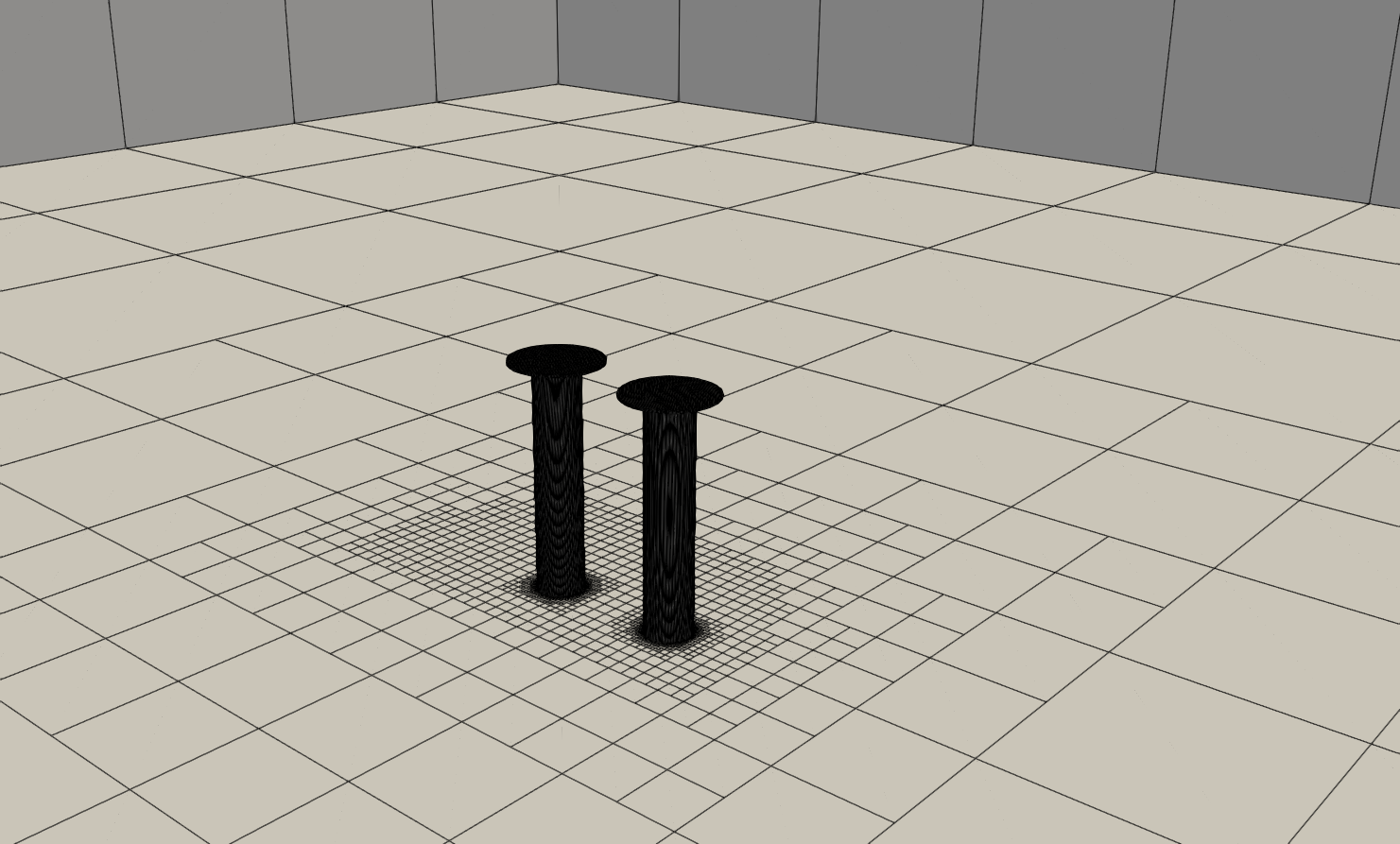}
    }
    \subfigure[]{
    \includegraphics[width=0.315\textwidth]{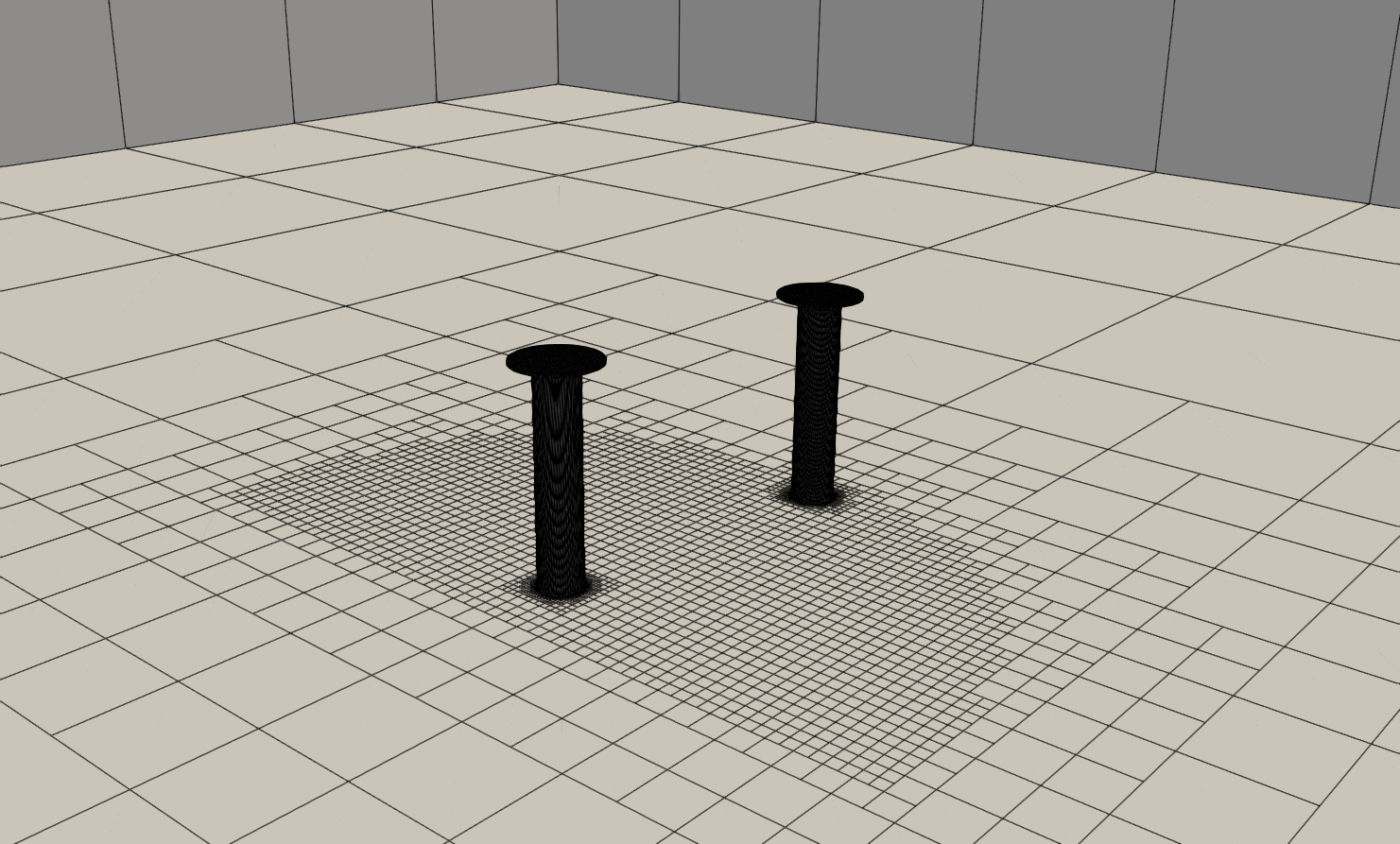}
    }
    \subfigure[]{
    \includegraphics[width=0.315\textwidth]{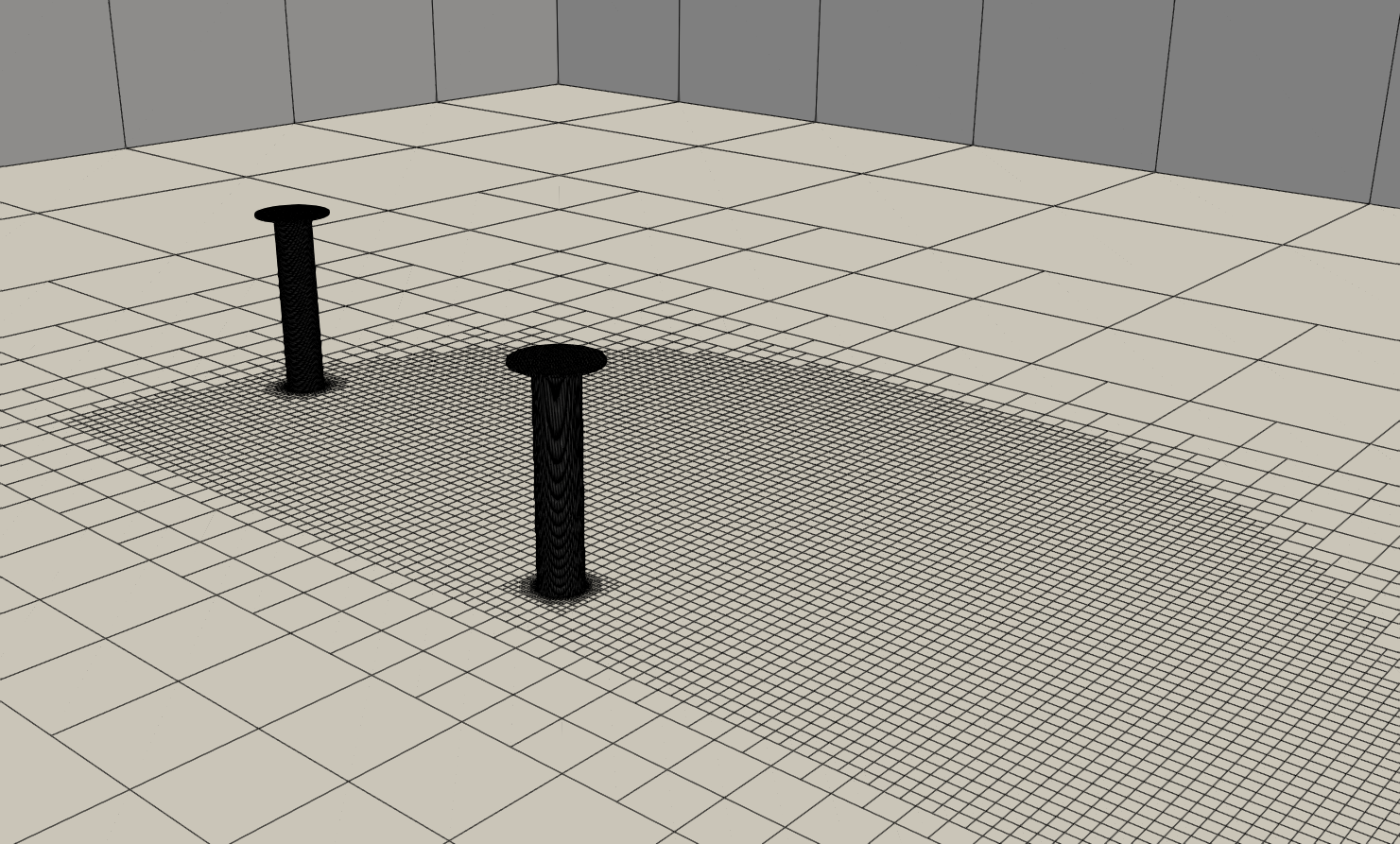}
    }
    \caption{Two interacting rotors case: Perspective view of the near-field grid for (a) configuration 1 ($\mathrm{WA} = 180^{\circ}$) at $\mathrm{GD} = 2D$ , (b) configuration 6 ($\mathrm{WA} = 90^\circ$) at $\mathrm{GD} = 6.5D$ and (c) configuration 11 ($\mathrm{WA} = 15^\circ$) at $\mathrm{GD} = 14D$.}
    \label{fig:interacting_rotors_grids}
\end{figure}

Findings and best practice settings of the previous investigations are utilized, e.g., Eulerian approaches utilize the UD ($\beta = 1$) approach. The computational domain extends over 100 rotor diameters in the longitudinal and lateral direction, with rotor A placed in the center, and 50 diameters in the vertical direction. The horizontal far fields correspond to wind boundaries; the pressure is specified at the upper domain end. Both rotors are positioned at the lower domain boundary, which is considered as a symmetry plane. The rotor surface is discretized homogeneously with approximately $\Delta x / D = 1/50$ elements, and the time step is selected so that an equivalent rotor flow is sampled with 100 time steps. The spatial range between the rotors is also refined with $\Delta x / D = 1/5$ to predict the interaction effect with sufficient accuracy. Hence, all resulting 39 Eulerian meshes feature around \SI{650000}{} control volumes, and expressions for selected configurations are provided in Fig. \ref{fig:interacting_rotors_grids}.

Benchmark results are generated using a URANS method for $\mathrm{Re_D} = 10^7$. For this purpose, a viscous layer with $y^+ \approx 50$ is added to all 39 Eulerian grids, which increases the number of control volumes sixfold to approximately \SI{3600000}{} per grid. Apart from that, the spatial and temporal discretizations are identical. In the case of the URANS method, the convective momentum transport is approximated using the TVD-QUICK method ($\kappa = 0.5$).
In all cases considered, the simulations are carried out for 50 equivalent flows, i.e., $T = 5000$ time steps, whereby the resulting forces are averaged over the last 10 flows.
As absolute comparisons have already been considered in Sec. \ref{subsec:single_rotor}, only relative differences are considered throughout this section.

The following Figs. \ref{fig:interacting_rotors_cd_rotor1}-\ref{fig:interacting_rotors_cl_rotor2} present relative differences in drag ($(c_\mathrm{d,A/B} - c_\mathrm{d,s.o.})/c_\mathrm{d,s.o.}$) and lift ($(c_\mathrm{l,A/B} - c_\mathrm{l,s.o.})/c_\mathrm{l,s.o.}$) coefficient of the respective interacting rotors against their stand-alone (s.o.) situation for all considered gap distances (blue: GD$=2$, red: GD$=6.5$ green: GD$=14$), at same spinning ratio, respectively. Therein, markings indicate the discrete sampling of the wind angle, whereby dashed lines with circles refer to inviscid and dashed lines with crosses to viscous studies. Rotor B's spinning ratio is increased from left to right, with each plot depicting all considered gap distances. 
Note that the consideration of rotor B's lift at $k_\mathrm{B} = 0$ utilizes an alternative normalization ($(c_\mathrm{l,B} - c_\mathrm{l,s.o.})/(c_\mathrm{l,s.o.} + 1)$) to avoid singularities due to vanishing lift coefficients.

Figures \ref{fig:interacting_rotors_cd_rotor1}-\ref{fig:interacting_rotors_cl_rotor1} present the effects of the non-repositioned rotor A's drag and lift coefficient interaction, respectively. In principle, the interactions are small for low rotational speeds of rotor B and increase as the rotational speed of rotor B increases. In addition, the interactions are more pronounced for small gap distances than in the case of more distant rotors. For an actively spinning rotor B, the drag coefficient of rotor A generally increases for positions in the wake of rotor A ($\mathrm{WA} < 90^\circ$) and decreases for positions in front of rotor A ($\mathrm{WA} > 90^\circ$). This trend has been predicted by both Eulerian and RANS methods, though the magnitudes and slope differ. On the other hand, the lift coefficient of rotor A tends to be 
increased in most cases at the presence of an actively spinning rotor B (the increase grows with higher spinning ratios of rotor B and short gap distance) and is only reduced for the extreme cases of rotor B positioned directly in front of or behind rotor A. 
\begin{figure}[!htb]
\centering
\iftoggle{tikzExternal}{
\input{./tikz/04__two_rotors/cd_rotor_1.tikz}}{
\includegraphics{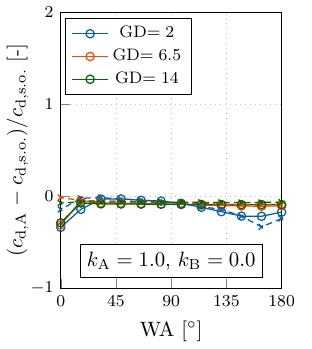}
\includegraphics{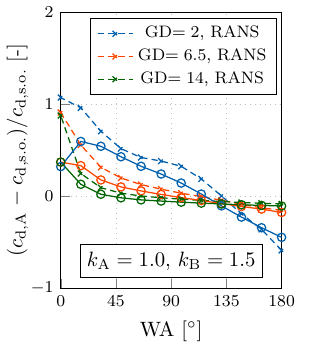}
\includegraphics{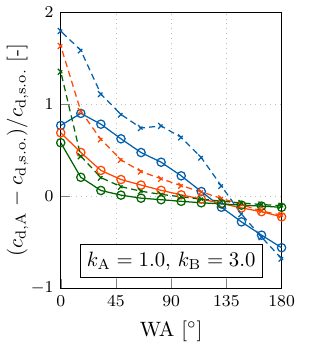}
}
\caption{Two interacting rotors case: Relative influence on the drag coefficient of the interacting rotor A ($c_\mathrm{d,A}$) against it's stand alone companion ($c_\mathrm{d,s.o.}$) with identical spinning ratio $k_\mathrm{A}=1$ for the Eulerian (EU) as well as a viscous (RANS) approach at varying Gap Distances (GD) per plot. Rotor B's spinning ratio varies from $k_\mathrm{B}=0$ (left) over $k_\mathrm{B}=1.5$ (center) to $k_\mathrm{B}=3$ (right).}
\label{fig:interacting_rotors_cd_rotor1}
\end{figure}
\begin{figure}[!htb]
\centering
\iftoggle{tikzExternal}{
\input{./tikz/04__two_rotors/cl_rotor_1.tikz}}{
\includegraphics{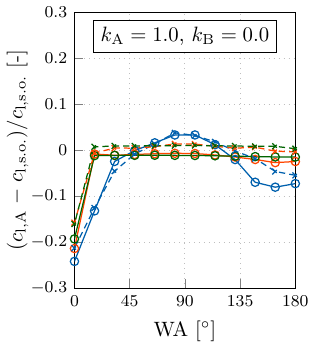}
\includegraphics{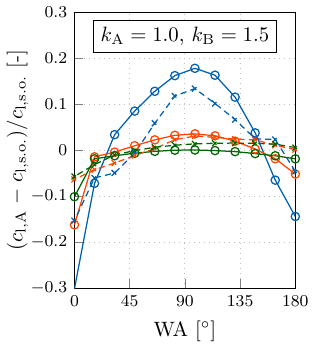}
\includegraphics{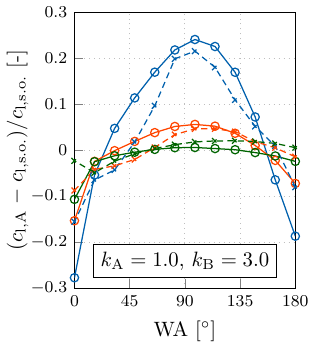}
}
\caption{Two interacting rotors case: Relative influence on the lift coefficient of the interacting rotor A ($c_\mathrm{l,A}$) against it's stand alone companion ($c_\mathrm{l,s.o.}$) with identical spinning ratio $k_\mathrm{A}=1$ for the Eulerian (EU) as well as a viscous (RANS) approach at varying Gap Distances (GD) per plot. Rotor B's spinning ratio varies from $k_\mathrm{B}=0$ (left) over $k_\mathrm{B}=1.5$ (center) to $k_\mathrm{B}=3$ (right).}
\label{fig:interacting_rotors_cl_rotor1}
\end{figure}

The interaction effects of rotor B are shown in Figs. \ref{fig:interacting_rotors_cd_rotor2}-\ref{fig:interacting_rotors_cl_rotor2}. The interaction of the drag coefficient shows an opposite trend over the wind angles compared to the graph of the drag coefficient of rotor A, particularly for $k_\mathrm{B} = [1.5,3]$ cf. Fig. \ref{fig:interacting_rotors_cd_rotor1} center and right. However, the intensity of the interaction is significantly reduced and amounts to values well below one. Once again, the interaction effects are most pronounced for rotors close to each other. This also applies to the interactions of the lift coefficient in Fig. \ref{fig:interacting_rotors_cl_rotor2}, whereby there is a tendency towards a reduction in rotor performance compared to the stand-alone situation, i.e., $(c_\mathrm{l,B} - c_\mathrm{l,s.o.})/c_\mathrm{l,s.o.} < 0$, which is in line with findings of \cite{kuhl2025adjoint} regarding the rotor-adjacent material positioning.

\begin{figure}[!htb]
\centering
\iftoggle{tikzExternal}{
\input{./tikz/04__two_rotors/cd_rotor_2.tikz}}{
\includegraphics{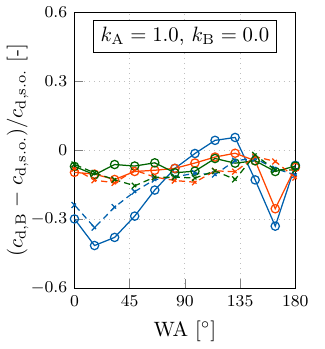}
\includegraphics{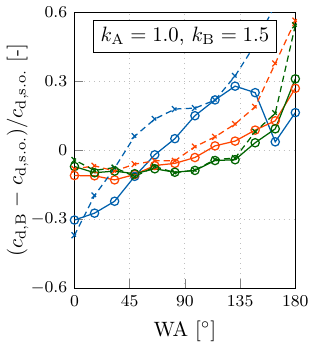}
\includegraphics{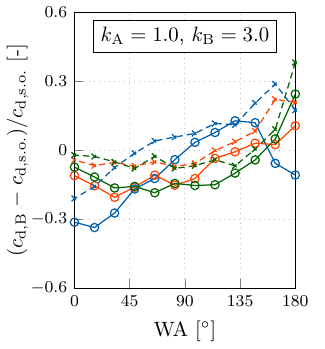}
}
\caption{Two interacting rotors case: Relative influence on the drag coefficient of the interacting rotor B ($c_\mathrm{d,B}$) against it's stand alone companion ($c_\mathrm{d,s.o.}$) with identical spinning ratios for the Eulerian (EU) as well as a viscous (RANS) approach at varying Gap Distances (GD) per plot. Rotor B's spinning ratio varies from $k_\mathrm{B}=0$ (left) over $k_\mathrm{B}=1.5$ (center) to $k_\mathrm{B}=3$ (right).}
\label{fig:interacting_rotors_cd_rotor2}
\end{figure}
\begin{figure}[!htb]
\centering
\iftoggle{tikzExternal}{
\input{./tikz/04__two_rotors/cl_rotor_2.tikz}}{
\includegraphics{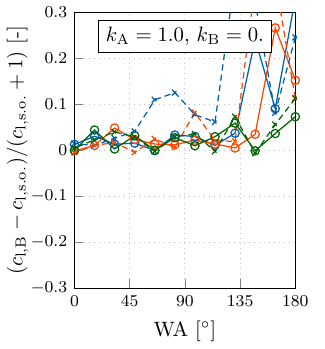}
\includegraphics{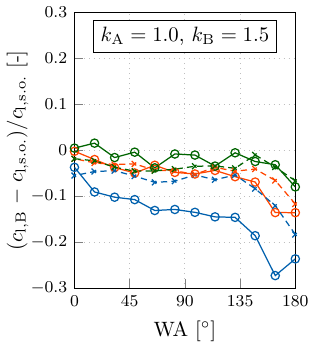}
\includegraphics{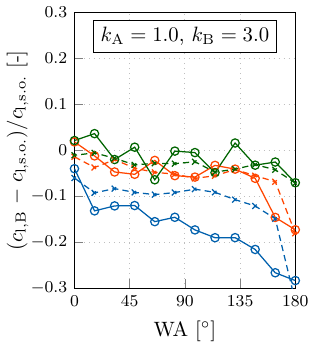}
}
\caption{Two interacting rotors case: Relative influence on the lift coefficient of the interacting rotor B ($c_\mathrm{l,B}$) against it's stand alone companion ($c_\mathrm{l,s.o.}$) with identical spinning ratios for the Eulerian (EU) as well as a viscous (RANS) approach at varying Gap Distances (GD) per plot. Rotor B's spinning ratio varies from $k_\mathrm{B}=0$ (left) over $k_\mathrm{B}=1.5$ (center) to $k_\mathrm{B}=3$ (right).}
\label{fig:interacting_rotors_cl_rotor2}
\end{figure}

The total simulation times are listed in Tab. \ref{tab:two_rotors_numerical_effort} according to the different modeling approaches. The numerical effort is measured in \#CPUh, follows from the number of \#CPUs used, the number of simulations performed (\#nSim) as well as the average simulation time ($\bar{t}^\mathrm{sim}$) in seconds, and is again significanlty reduced for the inviscid investigations.
\begin{table}[!ht]
\caption{Two interacting rotors case: Discretization and simulation data of both utilized simulation methods as well as resulting, overall required numerical effort.}
\begin{center}
\begin{tabular}{lcccccc}
\toprule
approach    & nCells [-]        & nTimeSteps [-]        & nCPU [-]          & nSim [-]      & $\bar{t}^\mathrm{sim}$ [s]        & CPUh [-]  \\ \midrule
URANS       & \SI{3672467}{}    & \SI{5000}{}           & 64                & 118           & \SI{0.320336E+05}{}               & \SI{66629}{}     \\
Eulerian    & \SI{640684}{}     & \SI{5000}{}           & 64                & 118           & \SI{0.837349E+03}{}               & \SI{1742}{}        \\
\bottomrule
\end{tabular}
\end{center}
\label{tab:two_rotors_numerical_effort}
\end{table}



\section{Application: Design Space Exploration for a Tanker Vessel}
\label{sec:application}
Building upon the validated framework introduced in the preceding sections, this final study demonstrates the applicability of the proposed numerical method to a realistic design task. Specifically, the workflow is applied to a parametric investigation of Flettner rotor installations on a representative tanker vessel. The objective is to demonstrate the method’s capability in supporting early-stage design decisions by enabling rapid and structured resolution of aerodynamic trends across a large design space.

\begin{figure}[!htb]
    \vspace*{\fill}
    \centering
    \subfigure[]{
    \includegraphics[width=0.6\textwidth,angle=0,origin=c]{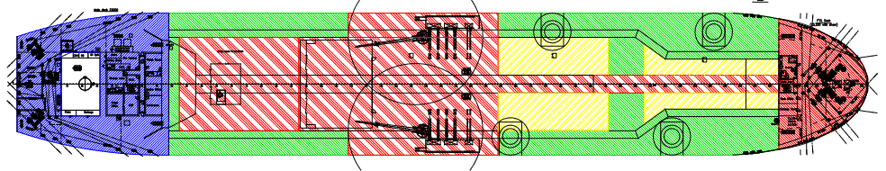}
    }
    \subfigure[]{
    \includegraphics[width=0.3\textwidth]{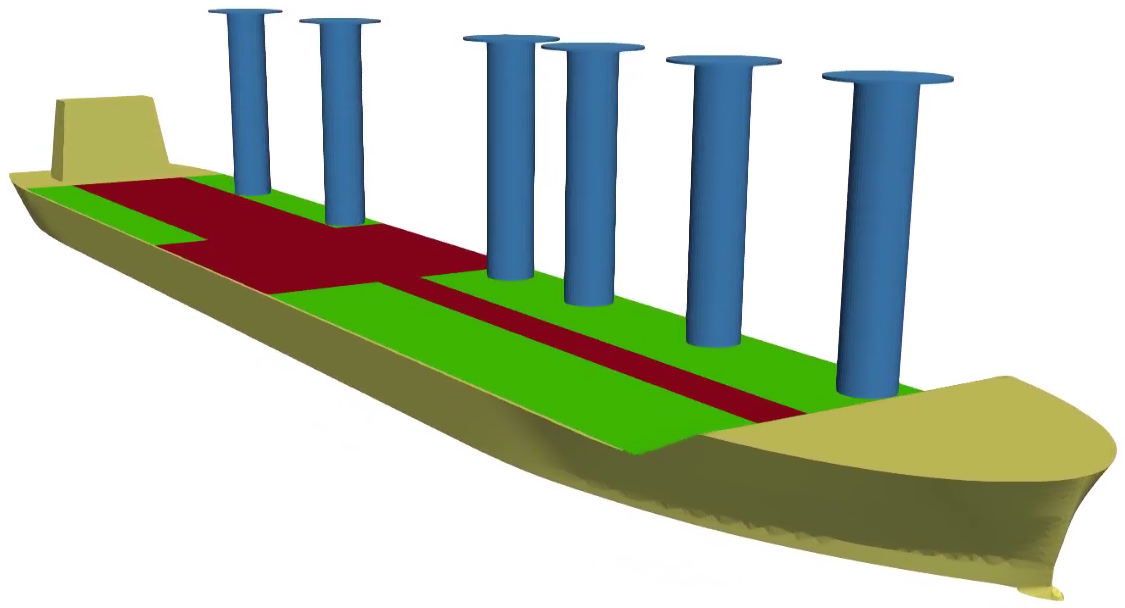}
    }
    \caption{Tanker design space exploration: General arrangement sketch (a) and simplified computational model with one considered rotor configuration (b). Green regions indicate regions allowed for rotor placement.}
    \label{fig:tanker_design_reion}
\end{figure}

The hull geometry is defined by the following principal characteristics: vessel length between the perpendiculars $L_\mathrm{pp} = \SI{224.9}{m}$, its beam $B = \SI{36.6}{m}$, and its design draft $T = \SI{12.8}{m}$. To approximate realistic onboard conditions, the vessel includes generic superstructures and simplified deck components for typical flow obstructions. The ship is considered above its waterline in line with the study from Sec. \ref{sec:annika_braren}. A large contiguous deck area amidships (colored green) remains unobstructed and serves as the layout surface for rotor installation, cf. Fig. \ref{fig:tanker_design_reion}. All considered rotors feature the same geometric characteristics, i.e., a diameter of $D = \SI{6}{m}$, a height of $H = \SI{36}{m}$, and thus an aspect ratio of $H/D = 6$. The upper end plate extends twice the rotor diameter. The investigated wind scenario follows a ship speed of $V_\mathrm{ship} = \SI{12}{kn}$ as well as a logarithmic wind profile with a reference speed of $V_\mathrm{wind} = \SI{10}{m/s}$ at a height of $h_\mathrm{wind} = \SI{10}{m}$ above the waterline combined with a Hellmann exponent of $\mathrm{mExp} = 0.11$, cf. Alg. \ref{alg:initial_flow_field}.

The simulation setup is adapted to reflect realistic application constraints, incorporating a simplified hull geometry, plausible operational conditions, and placement restrictions due to deck structures. The resulting design space exploration offers insight into the sensitivity of aerodynamic performance metrics --such as net lift force and effective thrust potential-- to both geometric design and environmental conditions. In this context, a clear distinction is made between \emph{configurations} and \emph{cases}:
\begin{itemize}
    \item \textbf{Configurations} refer to distinct rotor layouts defined by the number ($N_\mathrm{rotor} \in \{2, 4, 6\}$, nR in Alg. \ref{alg:flow_solver}), placement, and geometry of the rotors. All rotors share identical dimensions and are placed in allowed regions, cf. green areas in general arrangement sketch in Fig. \ref{fig:tanker_design_reion} (a). Each configuration requires an independent computational mesh, which is generated automatically via a meshing pipeline.
    \item \textbf{Cases} denote the specific aerodynamic conditions under which a given configuration is evaluated. These include variations in true wind angle (sampled in $15^\circ$ increments across the operating envelope) and rotor spining ratio, which is uniformly set for all rotors and varied across the range $\lambda \in [0, 4]$. These cases are handled through the presented generic implementation of the rotor body force model from Algs. \ref{alg:initial_flow_field} - \ref{alg:body_force_update}, enabling flexible assignment of circulation levels and directions without modifying the solver setup.
\end{itemize}
The complete set of cases is systematically evaluated for every configuration under consideration, resulting in a combinatorial expansion of simulation runs. Consequently, the total computational workload scales proportionally with the product $n_\mathrm{config} \times n_\mathrm{case}$. Herein, the number of cases follows twenty-four considered true wind angles and five considered spinning ratios each, i.e., $n_\mathrm{case} = 24 \times 5 = 120$ that are investigated for $n_\mathrm{config} = 40$ configurations, resulting in a total amount of $n_\mathrm{config} \times n_\mathrm{case} = 40 \times 120 = 4800$ numerical investigations. This highlights the necessity for a highly efficient simulation approach, as provided by the inviscid formulation presented in this manuscript.

\subsection{Input-Output Power Balance and System Efficiency}
\label{subsec:rotor_efficiency}
To assess the overall performance of the rotor system, the input mechanical power $P_\mathrm{in}$ required to drive the rotors is compared against the useful output power $P_\mathrm{out}$ generated by the rotor thrust. The output power can be estimated from the thrust force produced by the rotors $F_\mathrm{thrust}$, and the ship velocity $V_\mathrm{ship}$, i.e., $P_\mathrm{out} = F_\mathrm{thrust} V_\mathrm{ship}$.
The input power has been estimated through an impirical approach, which will be explained below.

To quantify the mechanical effort required to operate the rotors, the input power $P_\mathrm{in}$ needed to overcome viscous friction on the rotor surface is estimated. This power represents the mechanical energy that must be supplied to maintain rotor rotation against aerodynamic resistances. The input power can be expressed as the product of the shear stress $\tau_\mathrm{w}$ on the rotor surface, the wetted surface area $A_\mathrm{w} \approx \pi D H$, and the circumferential velocity $V = v_i t_i = \omega R$ of the rotor surface, i.e.,
\begin{align}
    P_\mathrm{in}
    \approx \tau_\mathrm{w} A_\mathrm{w} V
    = \tau_\mathrm{w} 2 \pi H \omega R^2
    \qquad \qquad \text{with} \qquad \qquad
    \tau_\mathrm{w} = c_\mathrm{f} \frac{1}{2} \rho (\omega R)^2 \, ,
\end{align}
where the shear stress $\tau_\mathrm{w}$ is given by the skin friction coefficient $c_\mathrm{f}$, which captures the frictional characteristics of the rotor surface dependent on flow conditions. The rotor's wetted surface area is approximated by the lateral surface area of a cylinder, neglecting thom disk contributions. Assuming a turbulent boundary layer on the rotor's smooth surface, the skin friction coefficient can be empirically approximated by the 1/7-power law (cf. \cite{schlichting2006grenzschicht}), viz., 
\begin{align}
    c_\mathrm{f} = 0.0576 \cdot \mathrm{Re}^{-1/5}
    \qquad \qquad \text{with} \qquad \qquad
    \mathrm{Re} = \frac{2 \omega R^2}{\nu} \, ,
\end{align}
where the kinematic viscosity of air $\nu$ is used to compile a Reynolds number, which characterizes the expected flow regime. As a result, the input power required to sustain rotor rotation against hypothetical, not resolved frictional resistance finally reads
\begin{align}
    P_\mathrm{in} = 0.0576 \left( \frac{ 2 \omega R^2}{\nu} \right)^{-1/5} \rho \pi H \omega^3 R^4
    \qquad \quad \to \qquad \quad
    P_\mathrm{in} = \alpha \, n^{2.8}
    \qquad \quad \text{with} \qquad  \quad
    \alpha = 0.0689 \, .
\end{align}
Based on the pursued approach, each rotor's input power is solely a function of the rotational rate scaled by a factor $\alpha = 0.0689$ based on the rotor and fluid-specific data. This relation provides a rough estimate of the mechanical power input, accounting for rotor geometry, rotational speed, and fluid properties, and serves as a practical model for preliminary design assessments. 

Finally, the system efficiency $\eta$ is then defined as the ratio of input to output power
\begin{align}
    \eta = \frac{P_\mathrm{in}}{P_\mathrm{out}} = \frac{P_\mathrm{in}}{F_\mathrm{thrust} V_\mathrm{ship}} \, .
\end{align}
This power balance provides a direct measure to evaluate the effectiveness of different rotor configurations and operating conditions. An efficiency $\eta < 1$ indicates that the rotor system provides a net propulsion benefit, delivering more useful power than the mechanical power invested in rotor rotation. Conversely, $\eta > 1$ implies that the system consumes more power than it produces in thrust.
Further influencing aspects, such as energy conversion or gearing losses, are not considered.



\subsection{Considered Design Space and Evaluation}
\begin{figure}[!htb]
    \vspace*{\fill}
    \centering
    \subfigure[]{
    \includegraphics[width=0.475\textwidth]{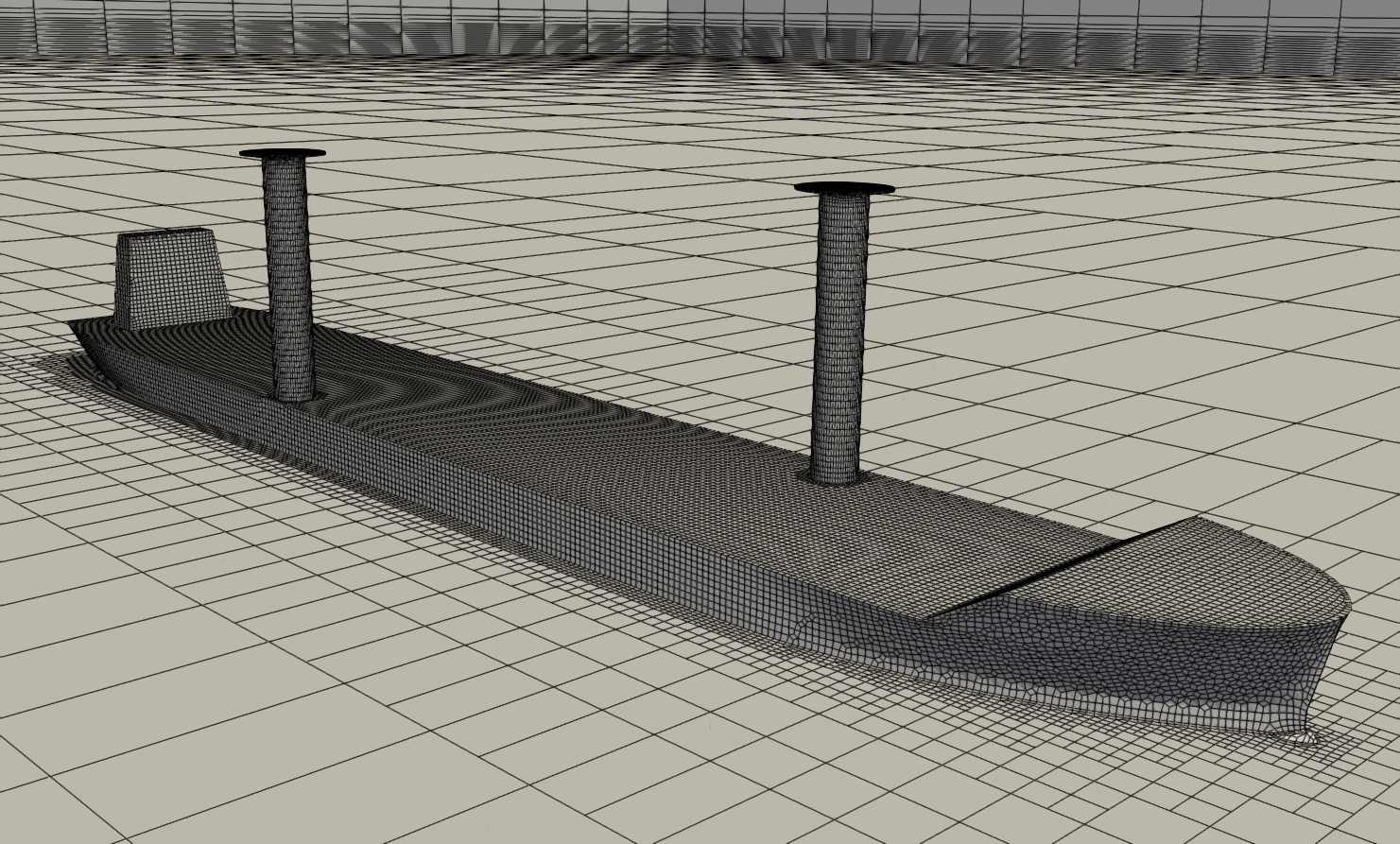}
    }
    \subfigure[]{
    \includegraphics[width=0.475\textwidth]{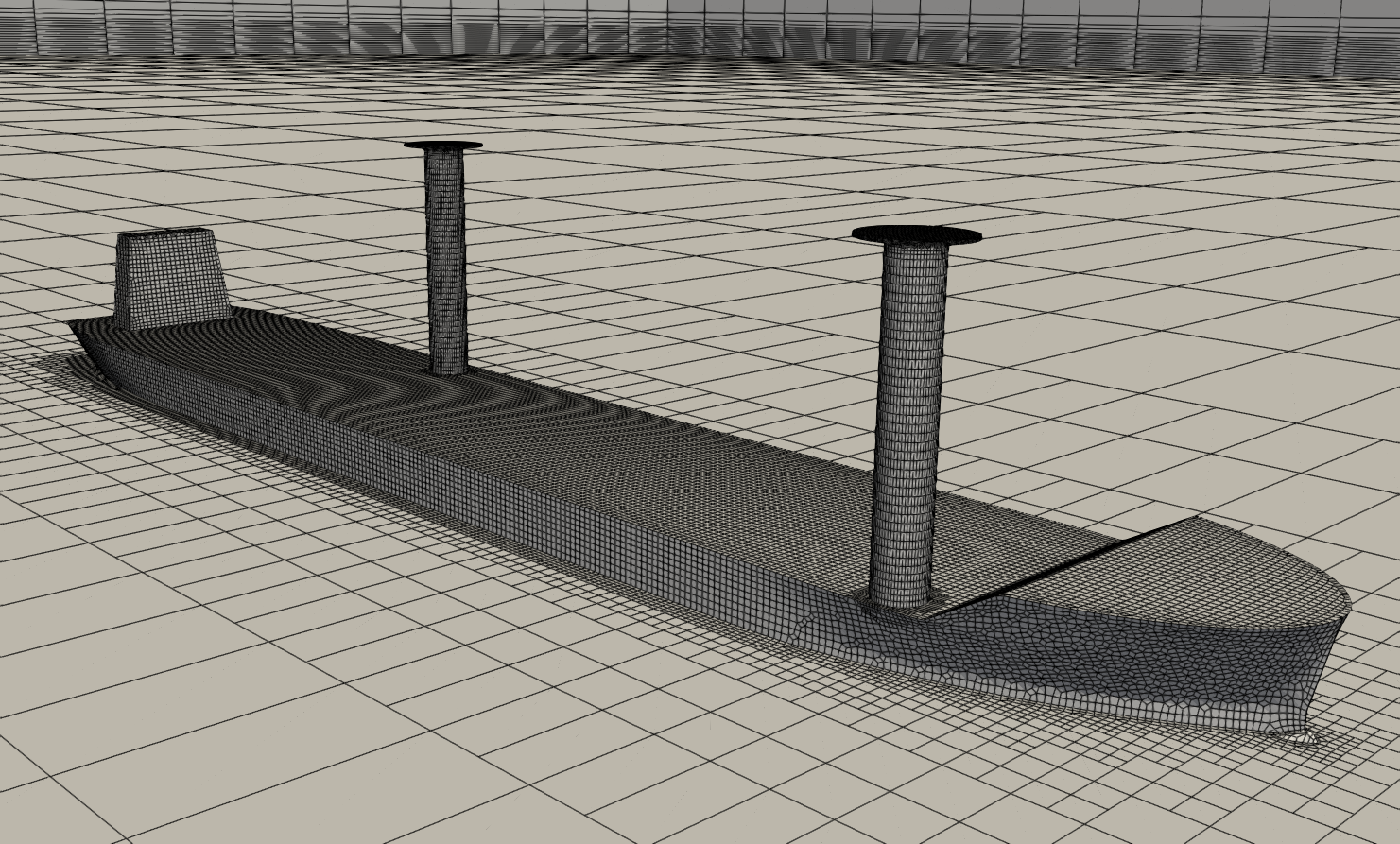}
    }
    \subfigure[]{
    \includegraphics[width=0.475\textwidth]{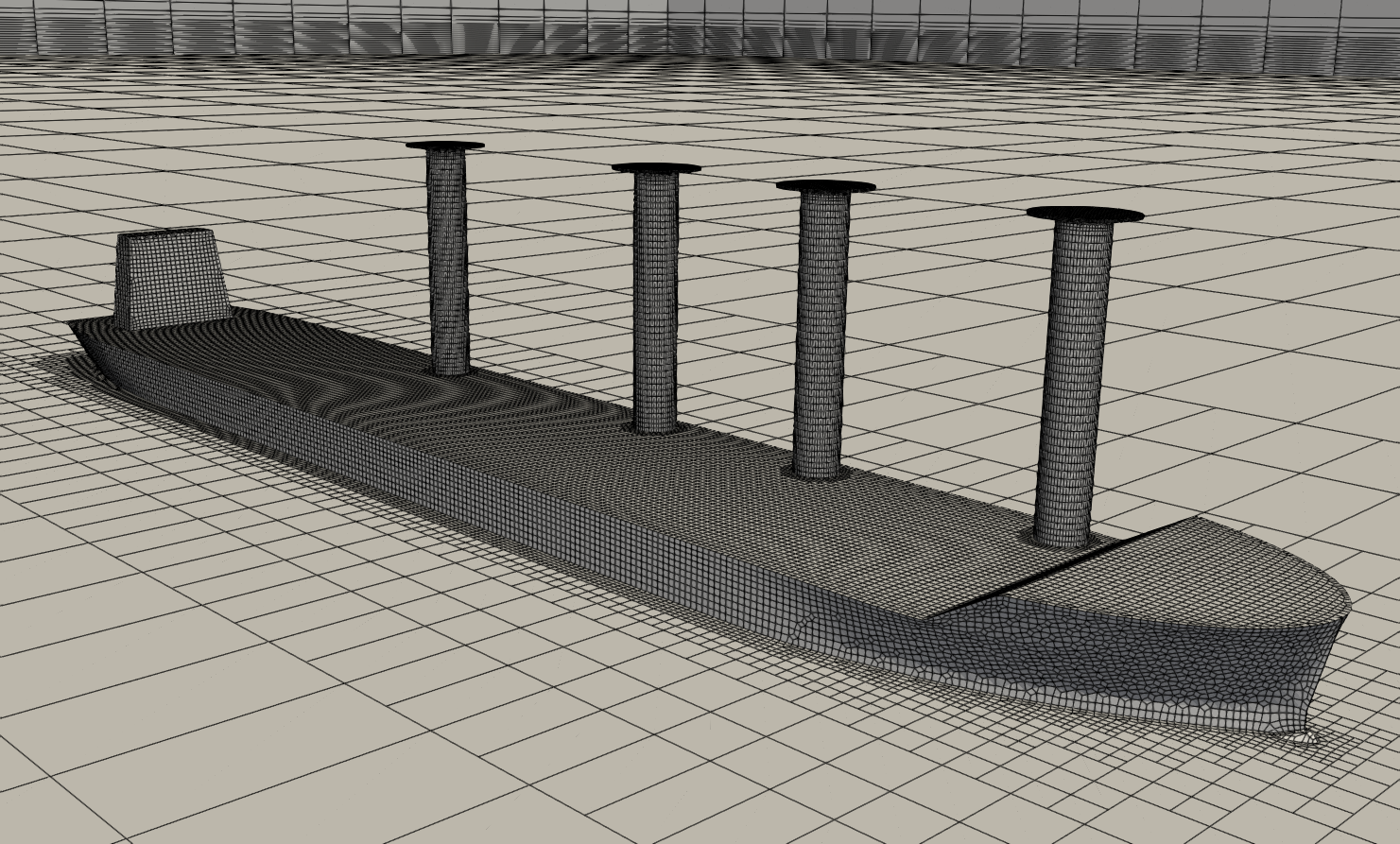}
    }
    \subfigure[]{
    \includegraphics[width=0.475\textwidth]{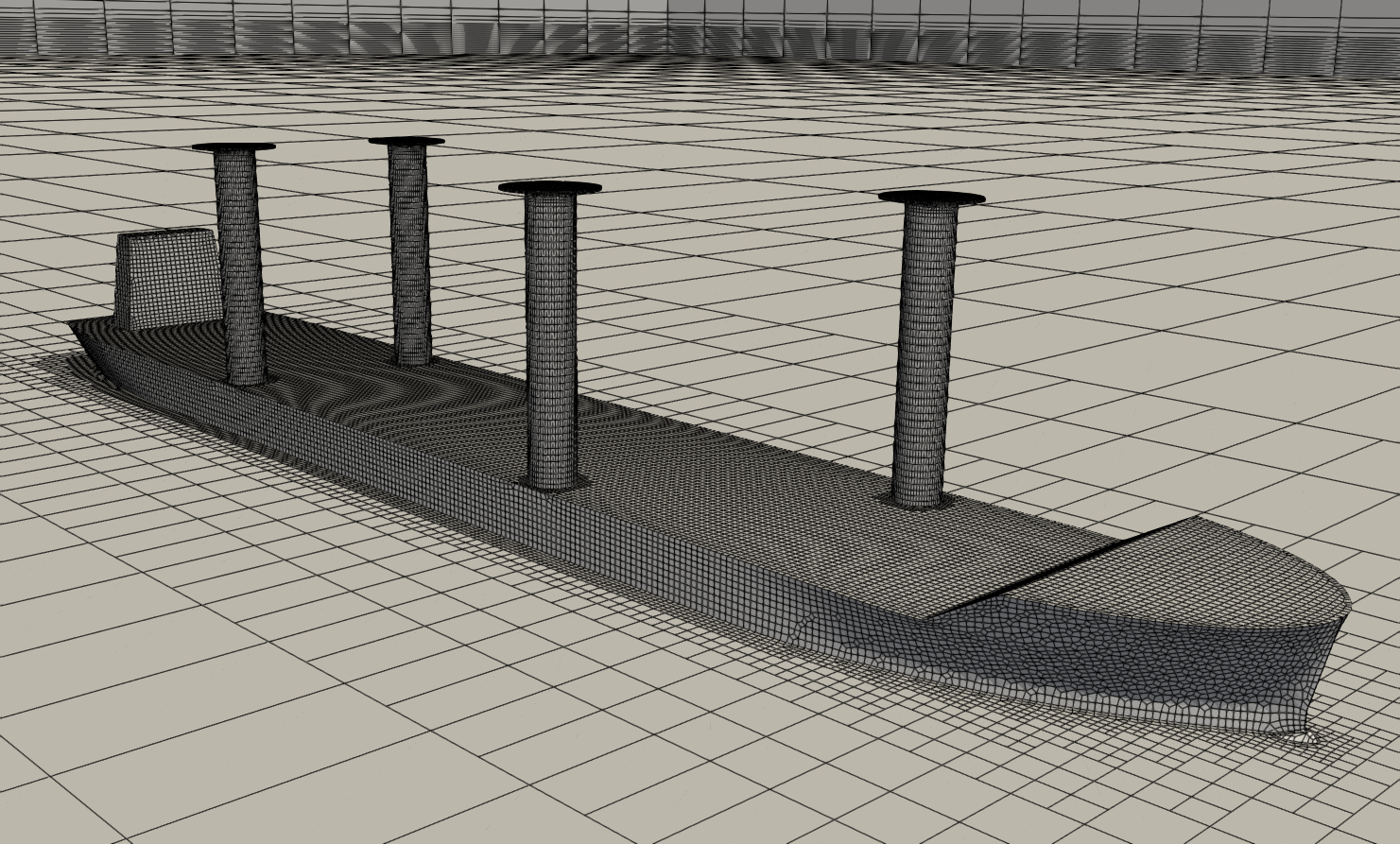}
    }
    \subfigure[]{
    \includegraphics[width=0.475\textwidth]{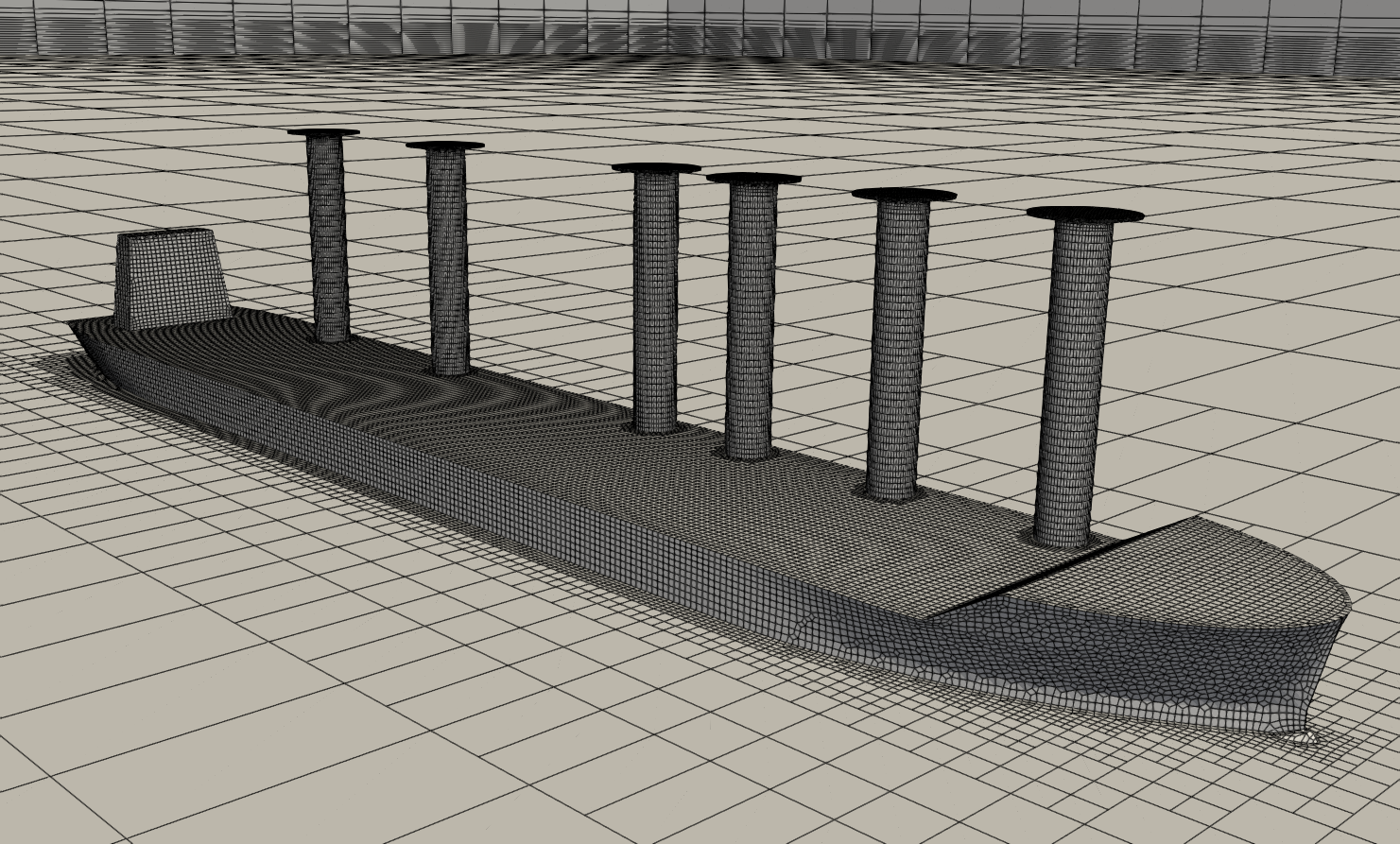}
    }
    \subfigure[]{
    \includegraphics[width=0.475\textwidth]{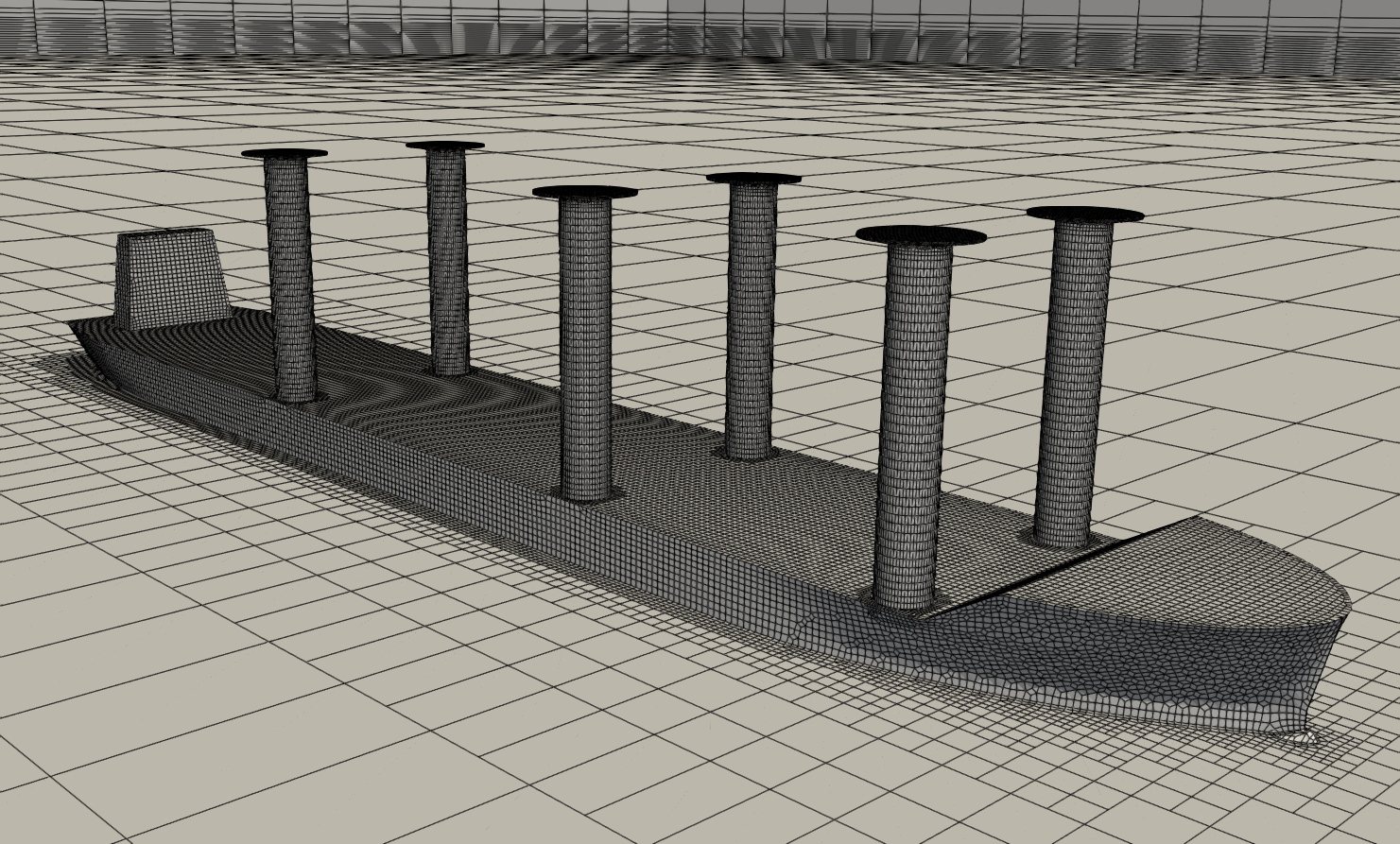}
    }
    \caption{Tanker design space exploration: Computational meshes of the best (left column) and second-best (right column) configuration for the considered two, four and six rotor studies, presented row wise from top to bottom, respectively.}
    \label{fig:tanker_best_configurations}
\end{figure}

A total of 40 variants were investigated, comprising 12 configurations for the 2-rotor case, 15 for the 4-rotor scenario, and 13 for the 6-rotor investigation. The configurations were designed to utilize the available installation space (see the green deck area in Fig. \ref{fig:tanker_design_reion}) as much as possible. For the investigations with 4 and 6 rotors, some positionings were also created randomly within constraints. Based on the maximum power output integrated across all wind angles and spinning ratios, the first and second-best configurations, as shown by their numerical grids in Fig. \ref{fig:tanker_best_configurations}, were identified. In the figure, the best configurations are shown in the left-hand column, and the second-best configurations are shown in the right-hand column. The number of rotors increases from top to bottom per row.

In the two-rotor case, the two best configurations tend to be symmetrical around the midship, with the slightly better variant positioning the front rotor on the port side and slightly more centrally than the second-best variant, which positions the front rotor on the starboard side and significantly further forward.

In the four- and six-rotor cases, the best configurations are those that position all rotors on the port side, essentially in a row. The second-best variants symmetrize the distributions again somewhat and assign the same number of rotors to both sides of the ship, whereby the six-rotor case results in an exact symmetry. The four-rotor case positions one of the two front rotors further forward and therefore adopts aspects of the outperforming two-rotor case. 

\subsection{Investigation of Best and Second-Best Configurations}
\begin{figure}[!htb]
    \centering
    \iftoggle{tikzExternal}{
    \input{./tikz/05__tanker/powerOut_3D.tikz}}{
    \includegraphics{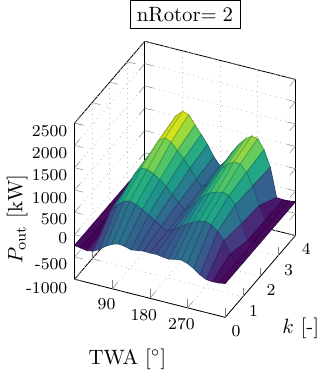}
    \includegraphics{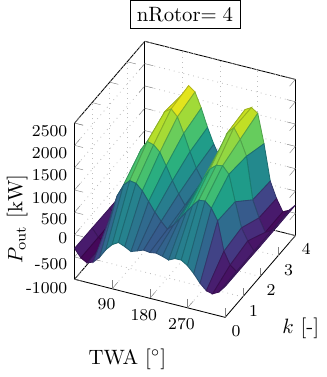}
    \includegraphics{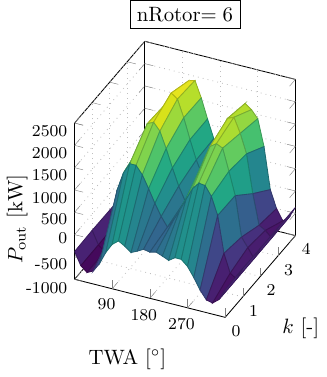}
    }
    \caption{Tanker design space exploration: Total power output of all rotors for the best two (left), four (center), and six (right) rotor cases over the investigated true wind angles and spinning ratios.}
    \label{fig:tanker_power_out_3D}
\end{figure}

\begin{figure}[!htb]
    \centering
    \iftoggle{tikzExternal}{
    \input{./tikz/05__tanker/powerOut_per_rotor_3D.tikz}}{
    \includegraphics{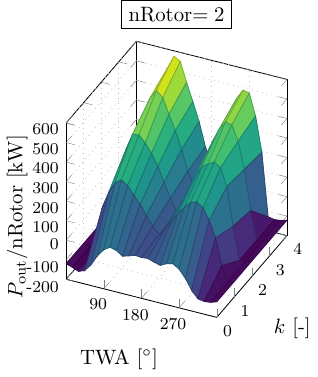}
    \includegraphics{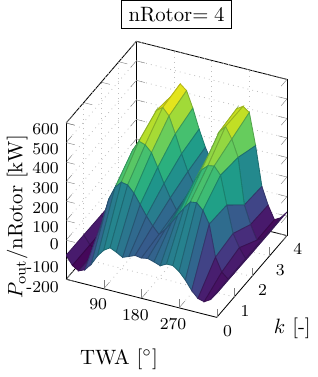}
    \includegraphics{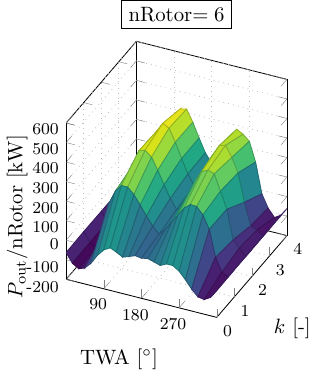}
    }
    \caption{Tanker design space exploration: Power output normalized by the number of rotors used for the best two (left), four (middle), and six (right) rotor cases over the true wind angles and spinning ratios investigated.}
    \label{fig:tanker_power_out_rotor_3D}
\end{figure}

The performance of the best and second-best configurations from Fig. \ref{fig:tanker_best_configurations} is examined in more detail below. First, the total power output in kW (see Sec. \ref{subsec:rotor_efficiency}) of the best two (left), four (center), and six (right) rotor cases are plotted in Fig. \ref{fig:tanker_power_out_3D} over the investigated wind angles and spinning ratios. Generally, the power output tends to increase for an increased number of rotors. Furthermore, the power output increases with the spinning ratio. It reaches a maximum for each spinning ratio at around 90 and 270 degrees, i.e., precisely when the wind hits the ship's side. The power output collapses for strong aft and forward winds. In the case of forward wind, the power output even becomes slightly negative, as the rotors generate resistance, even if they are possibly switched off.
The number of rotors normalizes the results in Fig. \ref{fig:tanker_power_out_3D} and the resulting power output per rotor is displayed in Fig. \ref{fig:tanker_power_out_rotor_3D}. Essentially, the impression from Fig. \ref{fig:tanker_power_out_3D} is inverted; i.e., studies with fewer rotors (left) deliver a higher power output per rotor than those with significantly more rotors (right). The ratio of the maxima is approximately two, i.e., with an actual wind angle of around 90 degrees and a spinning ratio of $k=4$, one rotor in the two-rotor scenario is twice as effective as one of the rotors in the six-rotor case.

\begin{figure}[!htb]
    \centering
    \iftoggle{tikzExternal}{
    \input{./tikz/05__tanker/powerOut.tikz}}{
    \includegraphics{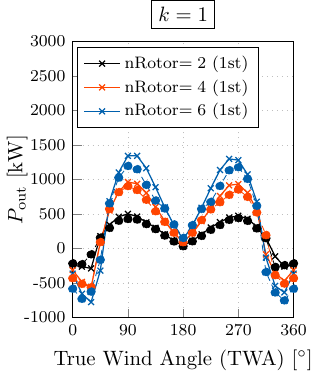}
    \includegraphics{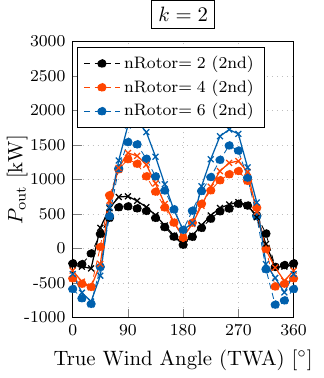}
    \includegraphics{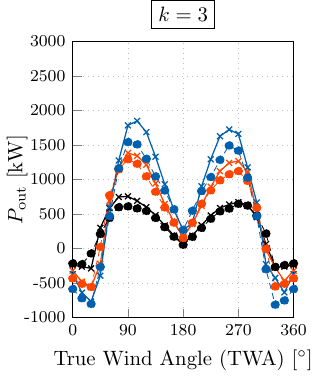}
    }
    \caption{Tanker design space exploration: Total power output of all rotors for the best (solid lines) and second-best (dashed lines) cases with two (black), four (orange), and six (blue) rotors above the true wind angle for a spinning ratio of $k=1$ (left), $k=2$ (center), and $k=3$ (right).}
    \label{fig:tanker_power_out_2D}
\end{figure}

Figure \ref{fig:tanker_power_out_2D} shows the power savings in two dimensions over the true wind angle for all finally identified configurations (two rotors: black, four rotors: orange, six rotors: blue), with the spinning ratio increasing from $k=1$ to $k=3$ from left to right. In addition to the best results (solid lines with cross markings), the results of the second-best configurations (dashed lines with filled circle markings) are now also shown. As this represents the total power output, the curves for the six-rotor cases are the largest, and those for the two-rotor cases are the smallest. In addition, the out-performance of the first placements compared to the second-best configurations is visible, with the differences being least pronounced for the small spinning ratio (left), especially in the two-rotor case.

\begin{figure}[!htb]
    \centering
    \iftoggle{tikzExternal}{
    \input{./tikz/05__tanker/powerOut_per_rotor.tikz}}{
    \includegraphics{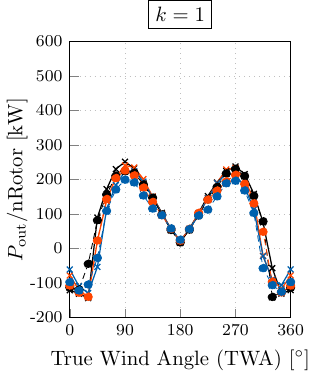}
    \includegraphics{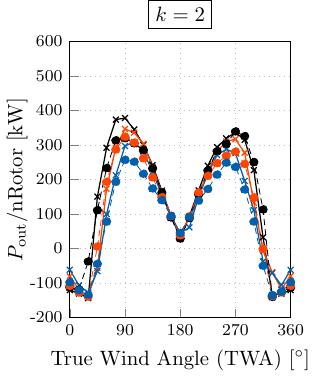}
    \includegraphics{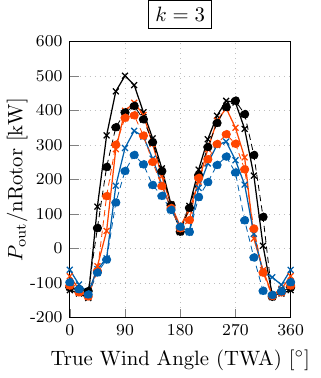}
    }
    \caption{Tanker design space exploration: Power output normalized by the number of rotors used for the best (solid lines) and second-best (dashed lines) cases with two (black), four (orange), and six (blue) rotors above the true wind angle for a spinning ratio of $k=1$ (left), $k=2$ (center), and $k=3$ (right).}
    \label{fig:tanker_power_out_rotor_2D}
\end{figure}

Analogous to Figs. \ref{fig:tanker_power_out_3D} - \ref{fig:tanker_power_out_rotor_3D}, the results from Fig. \ref{fig:tanker_power_out_2D} are now also normalized by the number of rotors, and the results are provided in Fig. \ref{fig:tanker_power_out_rotor_2D}, where the spinning ratio is increased from $k=1$ (left) to $k=2$ (center) and then to $k=3$ (right). The trend reversal already observed can be seen again, i.e., the studies with fewer rotors show a higher power output per rotor. As expected, the power output per rotor is lowest for low spinning ratios and higher for rotational speeds. Additionally, the out-performance of the first-place winner compared to the second-place winner persists.

\begin{figure}[!htb]
    \centering
    \iftoggle{tikzExternal}{
    \input{./tikz/05__tanker/ratio.tikz}}{
    \includegraphics{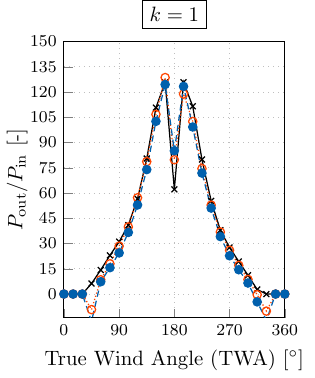}
    \includegraphics{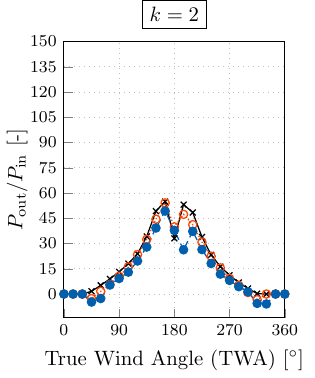}
    \includegraphics{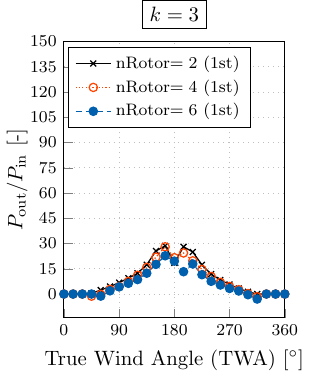}
    }
    \caption{Tanker design space exploration: Rotor efficiency of the best two (black), four (orange), and six (blue) rotor configurations above the true wind angle for spinning ratios of $k=2$ (left), $k=3$ (center), and $k=4$ (right).}
    \label{fig:tanker_rotor_efficiency}
\end{figure}

Finally, the rotor efficiency from Sec. \ref{subsec:rotor_efficiency} is examined, which is shown in Fig. \ref{fig:tanker_rotor_efficiency} for each of the best rotor configurations (two rotors: solid black, four rotors: dotted orange, six rotors: dashed blue) over the true wind angle, with the spinning ratio varying from $k=2$ (left) to $k=3$ (center) to $k=4$ (right). The rotor efficiency decreases significantly with increasing spinning ratios but remains well above 1 in almost all cases. It should be noted that the rotor is deactivated as soon as the power output becomes negative (rotor generates resistance only), cf. Alg. \ref{alg:body_force_update}.

\subsection{Rotor Interaction and Input for Power Prediction Tools}
In the context of ship routing and performance optimization, power prediction tools aim to integrate all relevant forces and moments acting on a vessel to anticipate its dynamic behavior and fuel demand under varying environmental conditions. When such tools include wind-assisted propulsion systems—such as Flettner rotors, it becomes essential to account for their aerodynamic contributions accurately, cf. \cite{xing2025wind}. To this end, normalizing the rotor’s lift and drag coefficients with respect to their standalone values provides a robust basis for quantifying interaction effects arising from the ship’s hull, deck structures, and local flow disturbances also due to presence of other rotors. This normalization enables the systematic incorporation of these effects into the force and moment balance used by the power prediction tool, thereby improving the reliability of route optimization and propulsion efficiency assessments.

The following Figs. \ref{fig:tanker_interaction_cd} - \ref{fig:tanker_interaction_cl} exemplary show normalized drag and lift coefficients of the rear three rotors of the best four-rotor case (cf. Fig. \ref{fig:tanker_best_configurations} (c)) over the true wind angle as well as the spinning ratio. The force coefficients from an idealized wind tunnel, e.g., from the investigations of Sec. \ref{subsec:single_rotor}, serve as reference values. A comparatively erratic picture can be observed, especially in the drag coefficient. In some cases, the drag coefficient even triples, illustrating the influence of the superstructure as well as the other rotors. However, note that the paper's inviscid flow formulation tends to overestimate induced resistances, cf., e.g., Fig. \ref{fig:annika_braren_cx_cy_TWA_grid2}. In addition, the interaction influences are more dependent on the wind angle than on the spinning ratio.
\begin{figure}[!htb]
    \centering
    \iftoggle{tikzExternal}{
    \input{./tikz/05__tanker/ecolibrium_2_rotors_fd.tikz}}{
    \includegraphics{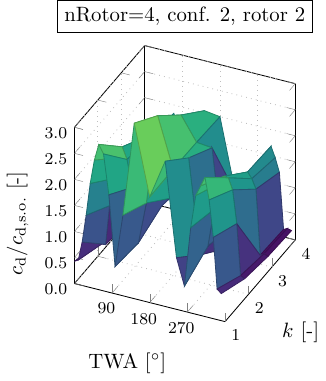}
    \includegraphics{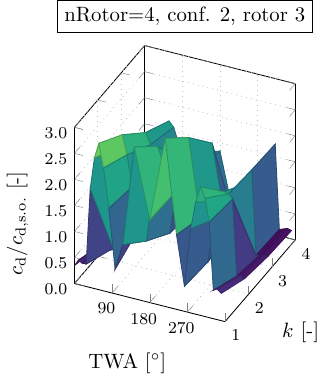}
    \includegraphics{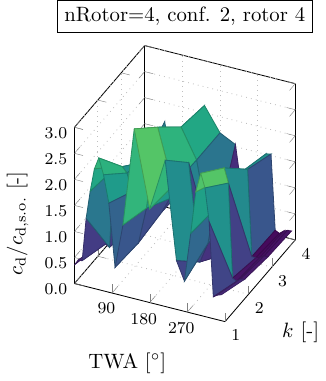}
    }
    \caption{Tanker design space exploration: Interaction factor between the drag coefficient of the second (left), third (center), and fourth (right) rotor of the best configuration with four rotors versus the idealized, stand-alone wind tunnel result over the true-wind angle and the spinning ratio.}
    \label{fig:tanker_interaction_cd}
\end{figure}

In the case of the lift coefficient, the interactions are not as pronounced in quantitative terms but are qualitatively similar to those of the drag coefficient. There is a tendency for a reduction in the lift coefficient compared to the idealized wind tunnel situation. For extreme headwind and tailwind situations, the lift coefficient drops sharply due to the deactivating of the rotors.
\begin{figure}[!htb]
    \centering
    \iftoggle{tikzExternal}{
    \input{./tikz/05__tanker/ecolibrium_2_rotors_fl.tikz}}{
    \includegraphics{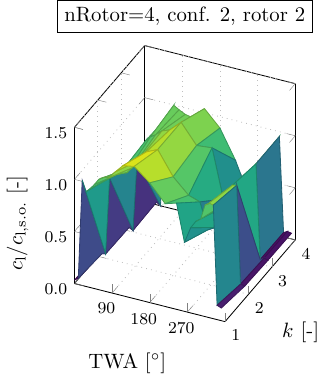}
    \includegraphics{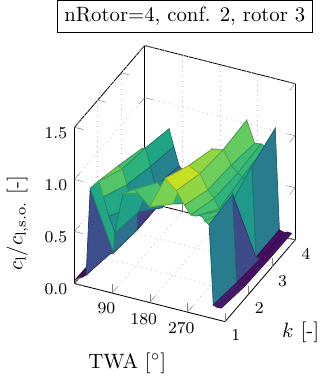}
    \includegraphics{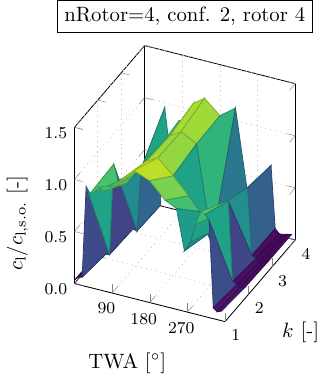}
    }
    \caption{Tanker design space exploration: Interaction factor between the lift coefficient of the second (left), third (center), and fourth (right) rotor of the best configuration with four rotors versus the idealized, stand-alone wind tunnel result over the true-wind angle and the spinning ratio.}
    \label{fig:tanker_interaction_cl}
\end{figure}

\begin{figure}[!htb]
    \centering
    \iftoggle{tikzExternal}{
    \input{./tikz/05__tanker/ecolibrium_2_rotors_onOff.tikz}}{
    \includegraphics{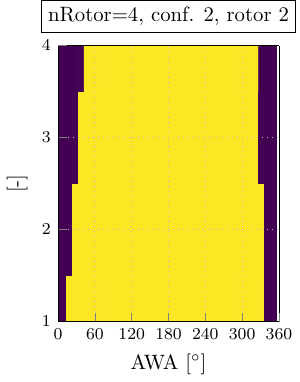}
    \includegraphics{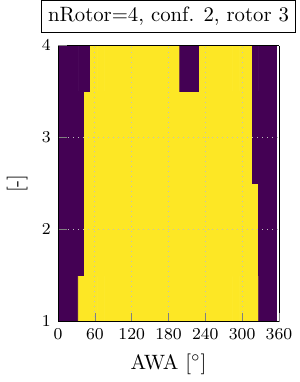}
    \includegraphics{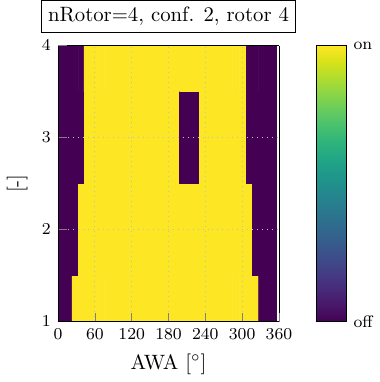}
    }
    \caption{Tanker design space exploration: Activity status of the first (left), second (center), and third (right) rotor of the best configuration with four rotors above the true-wind angle as well as the spinning ratio.}
    \label{fig:tanker_interaction_on_off}
\end{figure}

For a performance prediction tool, it is essential to know whether a rotor is switched on or off. Figure \ref{fig:tanker_interaction_on_off} shows the result from Alg. \ref{alg:body_force_update} as an example for the three rotors already discussed. It can be seen that the rotors are switched off in situations with a solid headwind and tailwind.

\section{Conclusion \& Outlook}
\label{sec:conclusion}

This work presented an inviscid CFD-based approach for rapid aerodynamic assessment of Flettner rotor installations, using a dynamic momentum-source formulation to enforce rotor circulation. The method enables efficient simulation of rotor-induced flow effects without requiring near-wall resolution or turbulence modeling, resulting in a substantial reduction in computational cost compared to viscous CFD approaches.

The validation studies demonstrate that the method is capable of capturing the dominant aerodynamic mechanisms, including lift generation, rotor–rotor, and rotor–superstructure interactions. While quantitative deviations from viscous reference solutions are observed, particularly at higher spinning ratios and depending on the numerical discretization, the approach consistently reproduces the main qualitative trends across the investigated parameter space.

These characteristics make the method particularly suitable for early-stage design applications, where large parameter spaces must be explored efficiently, and relative performance differences are of primary interest. In this context, the approach provides a robust and computationally efficient screening tool for the preliminary assessment and ranking of Flettner rotor configurations. At the same time, high-fidelity viscous simulations remain essential for final design validation.

Future work should further improve quantitative accuracy, for example, through hybrid approaches or calibration strategies, and extend the validation basis by incorporating additional experimental data and complex real-world configurations.

\section{Acknowledgments}
The current work is part of the research project “Development of a Comprehensive Methodology for the Integration of Flettner Rotors on Different Ship Types” (original German title: "Entwicklung der kombinierten aero- und hydrodynamischen Grundlagen für den Entwurf und den späteren optimierten Betrieb von Schiffen mit Windzusatzantrieben“), which is funded by the German Federal Ministry for Economic Affairs and Climate Action (Grant No. 03SX581G; Acronym FlettnerFleet). The authors gratefully acknowledge this support as well as the support from partners in this research project. We also acknowledge the helpful online guide provided by the University of Strathclyde under the umbrella of "Future Shipping: Flettner Rotors as Sustainable Propulsion" project website, which supported the rotor input power considerations discussed in Sec. \ref{sec:application}.

\section*{CRediT Authorship Contribution Statement}
\textbf{N.K.}: Conceptualization, Formal analysis, Investigation, Methodology, Software, Validation, Visualization, Writing – original draft, Writing – review \& editing
\textbf{Y.X.K.}: Formal analysis, Investigation, Methodology, Validation, Project administration, Writing - review \& editing

\section*{Declaration of Competing Interest}
The authors declare that they have no known competing financial interests or personal relationships that could have appeared to influence the work reported in this paper.

\section{Data Availability Statement}
Data sharing is not applicable to this article as no new data were created or analyzed in this study.

\end{document}